\documentclass[12pt]{article}

\pdfoutput=1
\usepackage{slashed}
\usepackage{color, verbatim}
\usepackage{latexsym}
\usepackage{epsfig}
\usepackage{amsmath,accents}

\usepackage{amssymb}
\usepackage{graphicx}
\usepackage{bm}

\usepackage[font={small}]{caption}

\usepackage{cite}
\usepackage{hyperref}

\newcommand{\hhref}[1]{\href{http://arxiv.org/abs/#1}{arXiv:#1}}

\definecolor{myred}{rgb}{0.7, 0, 0}
\definecolor{myblue}{rgb}{0, 0, 0.7}
\definecolor{mygreen}{rgb}{0.04, 0.7, 0.5}
\hypersetup{colorlinks,citecolor=myred,linkcolor=myblue,urlcolor=myblue,linktocpage=true}

\makeatletter

\@addtoreset{equation}{section}
\makeatother

\def\lsim{\mathrel{\raise.3ex\hbox{$<$\kern-.75em\lower1ex\hbox{$\sim$}}}}
\def\gsim{\mathrel{\raise.3ex\hbox{$>$\kern-.75em\lower1ex\hbox{$\sim$}}}}

\newcommand{\be}{\begin{equation}}
\newcommand{\ee}{\end{equation}}
\newcommand{\bea}{\begin{eqnarray}}
\newcommand{\eea}{\end{eqnarray}}

\newcommand{\tr}{\operatorname{tr}}
\newcommand{\diag}{\operatorname{diag}}

\begin{document}

\thispagestyle{empty}

\begin{center}

\hfill FERMILAB-PUB-18-386-T \\ 
\hfill EFI-18-9\\
\hfill UAB-FT-777

\begin{center}


{\Large\sc $R_{D^{(\ast)}}$ in custodial warped space}

\end{center}

\vspace{.5cm}

\textbf{
 Marcela Carena$^{\,a\,,b}$, Eugenio Meg\'ias$^{\,c}$, Mariano Quir\'os$^{\,d,\,e}$, Carlos Wagner$^{b\,,f}$
}\\

\vspace{1.cm}

${}^a\!\!$ {\em {
Fermi National Accelerator Laboratory, P.~O.~Box 500, Batavia, IL 60510, USA}}

${}^b\!\!$ {\em {
Enrico Fermi Institute and Kavli Institute for Cosmological Physics, \\ University of Chicago, Chicago, IL 60637, USA}}

${}^c\!\!$ {\em {Departamento de F\'{\i}sica At\'omica, Molecular y Nuclear and \\ Instituto Carlos I de F\'{\i}sica Te\'orica y Computacional, Universidad de Granada,\\ Avenida de Fuente Nueva s/n,  18071 Granada, Spain}}

${}^d\!\!$ {\em {Institut de F\'{\i}sica d'Altes Energies (IFAE) and BIST\\ Campus UAB, 08193, Bellaterra, Barcelona, Spain
}}

${}^e\!\!$ {\em {Department of Physics, University of Notre Dame\\ 225 Nieuwland Hall, Notre Dame, IN 46556, USA
}}

${}^f\!\!$ {\em {HEP Division, Argonne National Laboratory,9700 Cass Ave., Argonne, IL 60439, USA}}

\end{center}

\vspace{0.8cm}

\centerline{\bf Abstract}
\vspace{2 mm}
\begin{quote}\small
Flavor physics experiments allow to probe the accuracy of the Standard Model (SM) description at low energies, and are sensitive to
new heavy gauge bosons that couple to quarks and leptons in a relevant way.  The apparent anomaly in the ratios of the decay of 
$B$-mesons into $D$-mesons and different lepton flavors, $R_{D^{(\ast)}} = \mathcal B(B \to D^{(\ast)} \tau \nu)/ \mathcal B(B \to D^{(\ast)} \ell \nu )$ is particularly intriguing, since
these decay processes occur at tree-level in the SM. Recently, it has been suggested that this anomaly may be explained by new gauge bosons
coupled to right-handed currents of quarks and leptons, involving  light right-handed neutrinos. In this work we present a well-motivated ultraviolet 
complete realization of this idea, embedding the SM in a warped space with an $SU(2)_L \otimes SU(2)_R \otimes U(1)_{B-L}$ bulk gauge symmetry.
Besides providing a solution to the hierarchy problem, we show that this model, which has an explicit custodial symmetry, can explain the
$R_{D^{(\ast)}}$ anomaly and at the same time allow for a solution to the $R_{K^{(\ast)}}$ anomalies, related to the decay of $B$-mesons into $K$-mesons and 
leptons, $R_{K^{(*)}} =  \mathcal B(B\to K^{(*)} \mu \mu)/ \mathcal B(B \to K^{(*)} e e)$.  In addition,  a model prediction is an anomalous value of the forward-backward asymmetry $A^b_{FB}$, driven by the $Z\bar b_R b_R$ coupling, in agreement with LEP data. 
\end{quote}
\vfill

\newpage

\tableofcontents

\newpage

 \section{Introduction}
 \label{introduction}
 
 The Standard Model (SM) of particle physics provides an excellent description of all observables measured at collider experiments.
 The discovery of the Higgs boson~\cite{Aad:2012tfa,Chatrchyan:2012xdj} is an evidence of the 
 realization of the simplest electroweak symmetry breaking
 mechanism, based on the vacuum expectation value (VEV) of a Higgs doublet. This mechanism provides a moderate
 breakdown of the custodial $SU(2)_R$ symmetry that affects the gauge bosons only at the loop level.  The
predictions of the SM are also in agreement with precision electroweak observables,  which show only loop-size
departures from the tree-level gauge predictions~\cite{Patrignani:2016xqp}. 

Flavor physics experiments allow to further probe the accuracy of the SM predictions. While studying  SM
rare processes, these experiments become sensitive to heavy new physics coupled in a relevant way to
quarks and leptons. Recently, the BABAR~\cite{Lees:2012xj,Lees:2013uzd}, BELLE~\cite{Huschle:2015rga,Sato:2016svk,Hirose:2016wfn,Abdesselam:2016cgx,Abdesselam:2016xqt} and LHCb~\cite{Aaij:2015yra} experiments have measured the ratio of 
the decay of $B$-mesons into $D$-mesons and different lepton flavors,
\begin{equation}
R_{D^{(*)}} = \frac{\mathcal B(B \to D^{(*)} \tau \nu)}{\mathcal B(B \to D^{(*)} \ell \nu)},\quad \ell=\mu,e .
\end{equation}
These decay processes occur at tree-level in the SM, and therefore can only be affected in a relevant
way by either light charged gauge bosons, or heavy ones strongly coupled to the SM fermion 
fields. Currently, the measurements of these experiments seem to suggest a deviation of a few tens
of percent from the SM predictions, a somewhat surprising result in view of the absence of any clear
LHC new physics signatures, or other similar deviations in  other flavor physics experiment.

In particular, the presence of new $SU(2)$ gauge interactions affecting the left-handed neutrinos, which
could provide an explanation of the new $R_{D^{(*)}}$ anomaly, is strongly restricted by the measurement of the
branching ratio of the decay of $B$-mesons into $K$-mesons plus invisible signatures by the BELLE collaboration~$\mathcal B ( B \to K \nu\nu)$~\cite{Lees:2013kla,Lutz:2013ftz,Grygier:2017tzo}. Recently, it was proposed that a possible way
of avoiding these constraints was to assume that the new gauge interactions were coupled to right-handed
currents and the neutrinos are therefore right handed neutrinos~\cite{Asadi:2018wea,Greljo:2018ogz}.  The right-handed neutral currents are then
affected by right-handed quark mixing angles that are not restricted by current measurements, and provide
the freedom to adjust the invisible decays to values consistent with current measurements. 

In this work, we propose a well-motivated, ultraviolet complete, realization of the new gauge interactions coupled
to the right handed currents, by embedding the SM in warped space,
with a bulk gauge symmetry $SU(2)_L \otimes SU(2)_R \otimes U(1)_{B-L}$~\cite{Mohapatra:1974hk,Mohapatra:1974gc,Senjanovic:1975rk,Agashe:2003zs}. This symmetry is broken
to $SU(2)_L \otimes U(1)_Y$ in the ultraviolet brane, implying the absence of charged, $W_R^{\pm}$, and neutral, $Z_R$, gauge
boson zero modes. \textit{Third generation} quark and leptons are localized in the infrared-brane,
where a Higgs bi-doublet provides the necessary breakdown of the SM gauge symmetry, giving masses to quarks and leptons. 
Although there have been previous works on the flavor structure of warped extra dimensions with a $SU(2)_L \otimes SU(2)_R \times U(1)_X$ bulk gauge symmetry 
(see, for example, Refs.~\cite{Blanke:2008zb,Blanke:2008yr}), those works put emphasis on rare Kaon and $B$-meson decays
unrelated to $R_{D^{(*)}}$, that will also be analyzed in our work whenever relevant. Moreover, in the context of warped extra-dimensions, there has also been a recent analysis in Ref.~\cite{Blanke:2018sro} where lepto-quarks are introduced, and general results in composite Higgs models in Ref.~\cite{Azatov:2018knx}.

In this work, similarly to the previous proposal by the authors of Refs.~\cite{Asadi:2018wea,Greljo:2018ogz},
the new $SU(2)_R$ gauge bosons provide an explanation of the $R_{D^{(*)}}$ anomaly, and the freedom in
the right-handed mixing angles allows to avoid the invisible $B$ decay and $B$-meson mixing
constraints. On the other hand,
our model depicts unique, attractive special features such as having
an explicit custodial symmetry that protects it from large deviations in precision 
electroweak observables, and providing a solution of the hierarchy problem through the usual 
warped space embedding.  
Finally, although it is not the main aim of this article, the left-handed KK gauge bosons may be used
to provide an explanation of the $R_{K^{*}}$ anomalies in the way proposed in Refs.~\cite{Megias:2016bde,Megias:2017ove,Megias:2017vdg}.

Our study is organized as follows. In Sec.~\ref{model} we present the model in some detail. In Sec.~\ref{sec:RD}
we explain the solution to the $R_{D^{(*)}}$ anomaly. In Sec.~\ref{sec:Constraints} we discuss the 
existing experimental constraints on this model. In Sec.~\ref{sec:Predictions} we study the predictions of our model, including the forward-backward bottom asymmetry, the invisible decay of $B$ mesons into $K$ mesons, and the $b\to s \ell\ell$ observables, including $R_{K^{(\ast)}}$.
Finally we reserve Sec.~\ref{conclusions} for our conclusions and App.~\ref{sec:modes} for some technical details on the KK modes.

\section{The model}
\label{model}
Our setup will be a five dimensional (5D) model with metric (with the mostly minus signs convention) $g_{\mu\nu}=\exp(-ky)\eta_{\mu\nu}$, $g_{55}=-1$, in proper coordinates, and two branes, at the ultraviolet (UV) $y=0$, and infrared (IR) $y=y_1$, regions, respectively~\cite{Randall:1999ee}. The parameter $k$, close to the Planck scale, is related to the Anti de Sitter (AdS$_5$) curvature, and $k y_1$ has to be fixed by the stabilizing Goldberger-Wise (GW) mechanism~\cite{Goldberger:1999uk} to a value of $\mathcal O(35)$, in order to solve the hierarchy problem.

The custodial model is based on the bulk gauge group~\cite{Mohapatra:1974hk,Mohapatra:1974gc,Senjanovic:1975rk,Agashe:2003zs}
\be
SU(3)_c\otimes SU(2)_L\otimes SU(2)_R\otimes U(1)_X ,
\ee
where $X\equiv B-L$,
with 5D gauge bosons $(\mathcal G,W_L,W_R,X)$, and 5D couplings $(g_c,g_L,g_R,g_X)$~\footnote{The 5D ($g_5$) and 4D ($g_4$) couplings are related by $g_4=g_5/\sqrt{y_1}$.}, respectively. 

The breaking $SU(2)_R\otimes U(1)_X \to U(1)_Y$, where $Y$ is the SM hypercharge with gauge boson $B$ and coupling $g_Y$, is done in the UV brane by boundary conditions. Therefore the gauge fields $(W_L^a,W_R^a,X)$ define $(W_L^{a},W_R^{1,2},B,Z_R)$, with (UV, IR) boundary conditions, as
\begin{align}
&W_L^a\ (a=1,2,3), & (+,+)\\
B&=\frac{g_X W_R^3+g_RX}{\sqrt{g_R^2+g_X^2}},&(+,+)\\
&W_R^{1,2}, & (-,+)\\
Z_R&=\frac{g_R W_R^3-g_X X}{\sqrt{g_R^2+g_X^2}}. &(-,+)
\end{align}
The $SU(2)_L\otimes SU(2)_R$ symmetry is unbroken in the IR brane, where all composite states are localized, such that the \textit{custodial} symmetry is exact. In App.~\ref{sec:modes} we present some technical details leading to the wave function, mass and coupling of the $n\,th$ KK modes for both $(+,+)$ and $(-,+)$ boundary conditions. It is shown there that the difference for the KK mode masses $m_n$, and couplings, is tiny for the different boundary conditions, $(+,+)$ and $(-,+)$, and different electroweak symmetry breaking masses, and we will neglect it throughout this paper. In particular we will use the notation $m_1$ for the first KK mode mass of the different 5D gauge bosons after electroweak breaking: $(W_L^\pm,W_R^\pm,Z_R,Z_L,A)$.

The covariant derivative for fermions is
\be
\slashed{D}=\slashed\partial-i\left[g_L\sum_{a=1}^3 \slashed{W}_L^{a} T_L^a+g_R\sum_{b=1}^2 \slashed{W}_R^{b} T_R^b+
g_Y \slashed{B}\,Y+g_{Z_R}\slashed{Z}_R\,Q_{Z_R} ,
\right]
\ee
where $g_Y$ and $g_{Z_R}$ are defined in terms of $g_R$ and $g_X$ as
\be
g_Y=\frac{g_Rg_X}{\sqrt{g_R^2+g_X^2}},\quad g_{Z_R}=\sqrt{g_R^2+g_X^2}\ ,
\ee
and the hypercharge $Y$ and the charge $Q_{Z_R}$ are defined by
\be
Y=T_R^3+Q_X,\quad Q_{Z_R}=\frac{g_R^2T_R^3-g_X^2 Q_X}{g_R^2+g_X^2}
\ee
with $Q_X=(B-L)/2$.

Electroweak symmetry breaking is triggered in the IR brane by the bulk Higgs bi-doublet
\be
\mathcal H=\left( \begin{array}{cc}
H^{0}_2 & H_1^+ \\
H_2^- & H_1^0
\end{array}\right),\quad Q_X=0
\ee
where the rows transform under $SU(2)_L$ and the columns under $SU(2)_R$. We will denote their VEVs as $\langle H_2^0\rangle\equiv v_2/\sqrt{2}$ and $\langle H_1^0\rangle \equiv v_1/\sqrt{2}$, so that we will introduce the angle $\beta$ as, $\cos\beta=v_1/v_H$ and $\sin\beta=v_2/v_H$, with $v_H=\sqrt{v_1^2+v_2^2}$. We will find it useful to add an extra Higgs bi-doublet 
\be
\Sigma=\left(\begin{array}{cc}\Sigma^-/\sqrt{2} & \Sigma^0\\
\Sigma^{--} & -\Sigma^-/\sqrt{2}
\end{array}
\right),\quad Q_X=-1
\ee
with $\langle \Sigma^0\rangle=v_\Sigma/\sqrt{2}$,  whose usefulness will be justified later on in this paper.

After electroweak breaking, and rotating to the gauge boson mass eigenstates, one can re-write the covariant derivative as
\begin{align}
\slashed{D}=&\slashed{\partial}-ig_L\left[\frac{1}{\sqrt{2}}\ \slashed{W}_L^{\pm}T_L^{\pm}+\frac{1}{\cos\theta_L}\slashed{Z}_L\left( T_L^3-\sin^2\theta_L \, Q \right)  \right]-i\, g_L \sin\theta_L \slashed{A} \,Q
\nonumber\\
- &ig_R\left[ \frac{1}{\sqrt{2}}\ \slashed{W}_R^\pm T_R^\pm+\frac{1}{\cos\theta_R}\slashed{Z}_R\left(T_R^3-\sin^2\theta_R Y  \right)\right]
\label{eq:covariant}
\end{align}
where $\theta_L\equiv\theta_W$ is the usual weak mixing angle, the gauge boson $Z_L^\mu\equiv Z^\mu$, and $\theta_R$ is defined as
\be
\cos\theta_R=\frac{g_R}{\sqrt{g_R^2+g_X^2}},\quad \sin\theta_R=\frac{g_X}{\sqrt{g_R^2+g_X^2}}
\ee
with $T_{L,R}^\pm\equiv T_{L,R}^1\pm i T_{L,R}^2$. Using $g_R$ and $g_Y$, with $g_R>g_Y$, as independent parameters we can write
\be
g_X=\frac{g_Yg_R}{\sqrt{g_R^2-g_Y^2}},\quad \sin\theta_R=\frac{g_Y}{g_R},\quad
\cos\theta_R=\frac{\sqrt{g_R^2-g_Y^2}}{g_R} \,.
\label{eq:angles}
\ee


As for fermions, left-handed (LH) ones are in $SU(2)_L$ bulk doublets as in the SM
\be
Q_{L}^i=\left(\begin{array}{c} u_{L}\\ d_L\end{array}\right)^i,\quad L_L^i=\left(\begin{array}{c} \nu_L\\ e_L\end{array}\right)^i
\ee
where the index $i$ runs over the three generations. On the other hand, as $SU(2)_R$ is a symmetry of the bulk, right-handed (RH) fermions should appear in doublets of $SU(2)_R$. However, as $SU(2)_R$ is broken by the orbifold conditions on the UV brane it means, for bulk right-handed fermions, that one component of the doublet must be even, under the orbifold $\mathbb Z_2$ parity, and has a zero mode, while the other component of the doublet must be odd, and thus without any zero mode. We thus have to double the SM right-handed fermions in the bulk. 

The natural assignment is to assume in the bulk first and second (light) generation fermions:
\be
U_{R}^I=\left(\begin{array}{c} u_R\\ \tilde d_R \end{array}\right)^I,\
D_{R}^I=\left(\begin{array}{c} \tilde u_R \\ d_R\end{array}\right)^I,\
E_{R}^I=\left(\begin{array}{c} \tilde \nu_R \\ e_R\end{array}\right)^I,\ (I=1,2)
\ee
where only the untilded fermions have zero modes, while third generation (heavy) fermions are localized on the IR brane and thus are in $SU(2)_R$ doublets as
\be
Q_{R}^3=\left(\begin{array}{c} t_R\\ b_R\end{array}\right), \quad
L_{R}^3=\left(\begin{array}{c} \nu_{R}\\ \tau_R\end{array}\right) \,.
\ee
Then only the third generation RH fermions interact in a significant way with the field $W_R$, and can give rise to a sizable $R_{D^{(\ast)}}$, as we will see.

We define the KK modes for gauge bosons as
\be
A_\mu(x,y)=\sum_{n=0}^\infty \frac{f_A^n(y)}{\sqrt{y_1}} A_\mu (x)
\label{eq:KKGB}
\ee
normalized as
\be
\int_0^{y_1}f_A^nf_A^mdy=y_1\delta_{nm}
\label{eq:normalizacion}
\ee
and such that the factor $1/\sqrt{y_1}$ in Eq.~(\ref{eq:KKGB}) is absorbed by the 5D gauge coupling in Eq.~(\ref{eq:covariant}) to become the corresponding 4D gauge coupling.
 Similar definitions hold for KK modes of $Z_L(x,y)$ and $W_L(x,y)$, while for KK modes of 
$Z_R(x,y)$ and $W_R(x,y)$ the sum extends from $n=1$.

From the covariant derivative (\ref{eq:covariant}) it is clear that the charged bosons $W_L^\pm$ only interact with left-handed fermions, while $W_R^\pm$ only interact with right handed fermions. The corresponding 4D Lagrangian can be written as
\be
\frac{g_L}{\sqrt{2}}\sum_{f_L,f_L^\prime}G^n_{f_Lf^\prime_L}\bar f_L \slashed{W}_L^n f^\prime_L+\frac{g_R}{\sqrt{2}}\sum_{f_R,f_R^\prime}G^n_{f_Rf^\prime_R}\bar f_R \slashed{W}_R^n f^\prime_R
\label{eq:lagch}
\ee
where, from now on, we are switching to the notation where $g_L$ and $g_R$ are the 4D couplings, and $G^n_{f_{L,R}f_{L,R}^\prime}$ are the overlapping integrals of the fermion zero-mode profiles, $f_{L,R}(y)f_{L,R}^\prime(y)$, with the gauge boson KK mode ones, $W^n_{L,R}(y)$. 
On the other hand, the neutral gauge bosons $A$, $Z_L$ and $Z_R$ interact with both chiralities, and we can thus define the 4D neutral current Lagrangian for KK modes as
\be
\mathcal L_n=\frac{g_L}{\cos\theta_L}\sum_f
G^n_f\bar f (g_{Z_Lff}\slashed{Z}_L^n+g_{Aff}\slashed{A}^n) f+ \frac{g_R}{\cos\theta_R}\sum_f
g_{Z_Rff}G^n_f\bar f \slashed{Z}_R^n f
\label{eq:lagn}
\ee
where for simplicity we have omitted the chirality indices and $G^n_f$ is the overlapping integral of zero modes fermion profiles, $f_{L,R}^2(y)$, with the one of the (neutral)
gauge boson KK modes. The 4D coupling of photons with fermions is defined as $g_{Aff}=\sin\theta_L\cos\theta_L \,Q$, the couplings of fermions with $Z_R$ are given by
\begin{align}
&g_{Z_Ru_Ru_R}=\frac{1}{2}-\frac{2}{3}\sin^2\theta_R &g_{Z_Ru_Lu_L}=-\frac{1}{6}\sin^2\theta_R \nonumber\\
&g_{Z_Rd_Rd_R}=-\frac{1}{2}+\frac{1}{3}\sin^2\theta_R  &g_{Z_Rd_Ld_L}=-\frac{1}{6}\sin^2\theta_R \nonumber \\
&g_{Z_R\nu_R\nu_R}=\frac{1}{2} &g_{Z_R\nu_L\nu_L}=\frac{1}{2}\sin^2\theta_R \nonumber\\
&g_{Z_Re_Re_R}=-\frac{1}{2}+\sin^2\theta_R  &g_{Z_Re_Le_L}=\frac{1}{2}\sin^2\theta_R\ ,
\label{eq:gZR}\\
\nonumber
\end{align}
and with $Z_L$ by
\begin{align}
&g_{Z_Lu_Ru_R}=-\frac{2}{3}\sin^2\theta_L &g_{Z_Lu_Lu_L}=\frac{1}{2}-\frac{2}{3}\sin^2\theta_L \nonumber\\
&g_{Z_Ld_Rd_R}=\frac{1}{3}\sin^2\theta_L  &g_{Z_Ld_Ld_L}=-\frac{1}{2}+\frac{1}{3}\sin^2\theta_L  \nonumber\\
&g_{Z_L\nu_R\nu_R}=0 & g_{Z_L\nu_L\nu_L}=\frac{1}{2}\hspace{2cm} \nonumber\\
&g_{Z_Le_Re_R}=\sin^2\theta_L & g_{Z_Le_Le_L}=-\frac{1}{2}+\sin^2\theta_L \ .
\label{eq:gZL}\\
\nonumber
\end{align}

The 5D Yukawa couplings for RH quarks localized on the IR brane are
\be
Y^{i3}_Q \bar Q_{L\,i}\mathcal H Q_{R\,3}+\widetilde Y^{i3}_Q \bar Q_{L\,i}\widetilde{\mathcal H} Q_{R\,3}+h.c.
\ee
where $\widetilde{\mathcal H}=i\sigma_2 \mathcal H^{\ast} i\sigma_2$, and for the bulk RH quarks
\be
Y^{iI}_Q\bar Q_{L\,i}\mathcal H U_{R_I}+\hat{Y}^{iI}_Q\bar Q_{L\,i}\mathcal H D_{R_I}+\widetilde Y^{iI}_Q\bar Q_{L\,i}\widetilde{\mathcal H} U_{R_I}+\hat{\widetilde{Y}}^{iI}_Q\bar Q_{L\,i}\widetilde{\mathcal H} D_{R_I}+h.c.
\ee
so that the 4D Yukawa matrices are given by
\begin{align}
Y^u_{iI}&=\left(\sin\beta\, Y_Q-\cos\beta \,\widetilde Y_Q\right)_{iI} F(c_{u_L^i},c_{u_R^I})\,,\nonumber\\
Y^u_{i3}&=\left(\sin\beta\, Y_Q-\cos\beta \,\widetilde Y_Q\right)_{i3} F_3(c_{u_L^i})\,,
\label{eq:Yu}
\end{align}
and
\begin{align}
Y^d_{iI}&=\left(\cos\beta\, \hat{Y}_Q-\sin\beta \,\hat{\widetilde{Y}}_Q\right)_{iI}F(c_{d_L^i},c_{d_R^I})\,, \nonumber\\
Y^d_{i3}&=\left(\cos\beta\, Y_Q-\sin\beta\, \widetilde Y_Q\right)_{i3}F_3(c_{d_L^i})\,.
\end{align}
In the previous expressions the 4D Yukawa matrices $Y^{u,d}_{ij}$ contain the 5D Yukawa matrices $Y_Q,\widetilde Y_Q, \hat{Y}_Q, \hat{\widetilde{Y}}_Q$ times the integrals overlapping the 5D profiles of the corresponding fermions with the profile of the Higgs acquiring vacuum expectation value, $h(y)\propto e^{\alpha ky}$,
\begin{align}
F(c_L,c_R)&=\frac{\sqrt{2(\alpha-1)(1-2c_L)(1-2 c_R)}}{\alpha-c_L-c_R} \nonumber \\
&\quad \times \frac{e^{(\alpha-c_L-c_R)ky_1} - 1}{\sqrt{\left[e^{2(\alpha-1) k y_1} -1\right] \left[e^{(1-2c_L)ky_1}-1 \right]\left[e^{(1-2c_R)ky_1}-1  \right]}} \,, \nonumber\\
F_3(c_L)&=\sqrt{2(\alpha-1)(1-2c_L)}\frac{e^{(\alpha-1/2-c_L)ky_1}}
{\sqrt{\left[e^{2(\alpha-1) k y_1} -1\right] \left[e^{(1-2c_L)ky_1}-1\right]}} \,,
\end{align}
where $c_{L,R}$ are the fermion bulk mass parameters and we have
assumed that $\alpha>c_L+c_R$. The parameter $\alpha$ has to be
larger than (or equal to) two, to solve the hierarchy problem, and in
our computations we will fix $\alpha=2$.

Similarly for RH leptons in the IR brane
\be
Y^{i3}_L\bar L_{L\,i}\mathcal H L_{R\,3}+\widetilde Y^{i3}_L\bar L_{L\,i}\widetilde{\mathcal H} L_{R\,3}+h.c.
\ee
and for bulk RH leptons
\be
Y_L^{iI}\bar L_{L}^i \mathcal H N_{R}^I+   \hat{Y}^{iI}_L\bar L_{L}^i\mathcal H E_{R}^I+
\widetilde{Y}^{iI}_L\bar L_{L}^i\widetilde{\mathcal H} N_{R}^I+
\hat{\widetilde{Y}}^{\prime\,iI}_L\bar L_{L}^i\widetilde{\mathcal H} E_{R}^I+h.c.
\ee
where we have added the bulk first and second generation right-handed neutrino doublets 
\be
N_R^I=\left(\begin{array}{c}\nu_R \\ e_R^\prime \end{array}
\right)^I,\ I=(1,2).
\label{eq:neur}
\ee

The Yukawa couplings for charged leptons are then given by
\begin{align}
Y^e_{i3}&=(\cos\beta Y_L-\sin\beta \widetilde Y_L)_{i3}F_3(c_{e_L^ i})\,,\nonumber\\
Y^e_{iI}&=(\cos\beta \hat{Y}_L-\sin\beta \hat{\widetilde{Y}}_L)_{iI}F(c_{e_L^i},c_{e_R^I}) \,,
\end{align}
and for neutrinos, by
\begin{align}
Y^\nu_{iI}&= (\sin\beta\, Y_L-\cos\beta\, \widetilde Y_L)_{iI}F(c_{\nu_L^i},c_{\nu_R^I}) \,, \nonumber\\
Y^\nu_{i3}&= (\sin\beta\, Y_L-\cos\beta\, \widetilde Y_L)_{i3}F_3(c_{\nu_L^i}) \,.
\label{Ynu}
\end{align}

In the presence of a non-zero vacuum expectation value of the $\Sigma$ field, we shall define
\be
\tan\theta_\Sigma = \frac{v_\Sigma}{v_H},
\ee
where $v = \sqrt{v_H^2 + v_\Sigma^2}$.  In the decoupling limit,
$H_1= \cos\theta_\Sigma \cos\beta h-\sin\beta H-\sin\theta_\Sigma \cos\beta H_\Sigma$  and $H_2=\cos\theta_\Sigma \sin\beta h+\cos\beta H-\sin\theta_\Sigma \sin\beta H_\Sigma$, while the neutral 
component of the $\Sigma$ field, $\Sigma^0 = \sin\theta_\Sigma h + \cos\theta_\Sigma H_\Sigma$. The SM-like Higgs 
boson is induced by excitations of the field $h = \sin\theta_\Sigma \Sigma^0 + \cos\theta_\Sigma ( \cos\beta H_1 + \sin\beta H_2)$,
while the excitations induced by the orthogonal combinations $H$ and $H_\Sigma$ are supposed to lead to heavy neutral states, decoupled
from the low energy theory. Since quarks and leptons only couple to the field $\mathcal H$, the masses are proportional to $v_H$ and therefore 
the Yukawa couplings must be enhanced by a factor $(\cos\theta_\Sigma)^{-1}$ with respect to the value they would obtain in the absence of the $\Sigma$ field.

In order to avoid strong constraints from lepton flavor violating processes, as e.g.~$\mu\to e\gamma$, $\mu\to eee$, or $\mu-e$ conversion, we will assume that for charged leptons the interaction and mass eigenstate bases coincide, and therefore, hereafter, that the matrix $Y^e$ is diagonal. 
This can be obtained by imposing a
$U(1)^3$ flavor symmetry in the lepton sector broken only by
the tiny effects due to the neutrino masses~\cite{Megias:2017isd}.

For neutrinos propagating  in the bulk, one can obtain realistic values of their masses 
 by adopting one of the proposed solutions for theories with warped extra dimensions~\cite{Huber:2003tu,Moreau:2005kz,Csaki:2008qq,Perez:2008ee,vonGersdorff:2012tt}. In our scenario, however, neutrinos localized on the IR brane, as is the case with the right-handed neutrinos $\nu_{\tau_R}$, couple in a relevant way to the Higgs and tend to acquire
 masses of the same order as the charged lepton masses. This can be seen from the fact that the Yukawa couplings in Eq.~(\ref{Ynu}) will provide a Dirac mass to the third generation neutrinos $m_D\bar\nu_L \nu_R+h.c.$. Therefore, in order to obtain realistic masses  
 we will assume a double seesaw scenario~\cite{Barr:2003nn}.  We shall first concentrate on the example of third generation neutrinos. In order to realize this mechanism, 
we will introduce a Higgs $H_R$, transforming as $(1,2,-1/2)$ under $SU(2)_L\otimes SU(2)_R\otimes U(1)_X$, which spontaneously breaks $SU(2)_R \times U(1)_X \to U(1)_Y$, when its neutral, hyperchargeless, component gets a vacuum expectation value $v_R$, as well as a localized fermion singlet $(1,1,0)$, $S_L$, which provides the Dirac mass $m_D^\prime \bar S_L \nu_R+h.c.$, where $m_D^\prime = Y_R v_R/\sqrt{2}$.  Finally, we can also write down a Majorana mass term as $M S_L S_L$. Therefore the mass matrix in the basis $(\nu_L,\nu_R^c,S_L)$ can be written as
\be
\mathcal M_\nu=\left(\begin{array}{ccc}0&m_D&0\\
m_D& 0&m_D^\prime \\
 0& m_D^\prime & M 
\end{array}
\right)  \,.
\label{eq:Mnu}
\ee
In the limit where $m_D^\prime\gg m_D\gg M$ there is a mass eigenstate $\nu_0\simeq \nu_L$ with a mass $m_{\nu_0}\simeq (m_D/m_D^\prime)^2 M$ (which is obviously massless in the limit where $M=0$), and an approximate Dirac spinor $\nu_1=(\nu_R^c-S_L,-\nu_R+S_L^c)^T/\sqrt{2}$, with a mass $m_{\nu_1}\simeq \sqrt{m_D^2+m_D^{\prime\,2}}$. This mechanism has been dubbed in the literature, double seesaw~\cite{Barr:2003nn}. The double seesaw mechanism allows for acceptable masses for the left- and right-handed
neutrinos without extreme fine-tuning of the Yukawa couplings. For instance, for $m_D \simeq 1$~MeV, $m_D^\prime \simeq 100$~MeV and $M = {\cal{O}}(1 \; {\rm KeV} )$, one 
obtains a mostly left-handed neutrino of mass of order 0.1~eV, and an additional pseudo-Dirac neutrino, containing $\nu_R$, of mass of order 100~MeV.  Such masses
are enough to accommodate the value of $R_{D^{(*)}}$ without any sizable kinematic suppression.

The above mechanism can be easily generalized to give mass to the three generations of  neutrinos. 
As suggested before, we will consider in the bulk the two RH neutrino doublets $N_R^I$ and add two singlets $S_L^I$, while the third
generation right-handed leptons and the singlet $S_L^3$ are as before localized in the IR brane. 
States transform under the flavor symmetry group $U(1)^3=U(1)_{L_e}\otimes U(1)_{L_\mu}\otimes U(1)_{L_\tau}$, where the lepton number is defined as
$L\equiv L_e+L_\mu+L_\tau$, in Tab.~\ref{tab:neutrinomasses}.
\begin{table}[htb]
\centering
\begin{tabular}{||c||c|c|c|c||}
\hline\hline
 & $L_e$ & $L_\mu$& $L_\tau$ & $L$ \\ \hline\hline
 $N_R^1$& 1 & 0 & 0 & 1 \\
 $N_R^2$& 0 & 1 & 0 & 1 \\
 $L_R^3$& 0 & 0 & 1 & 1 \\ \hline
 $H_R$ &1/3 &1/3 &1/3 &1   \\ \hline
 $S_L^1$ & 2/3 &-1/3  &-1/3  & 0   \\
  $S_L^2$ & -1/3 &2/3  &-1/3  & 0   \\
   $S_L^3$ & -1/3 &-1/3  &2/3  & 0   \\
\hline\hline
\end{tabular}
\caption{\it Leptonic quantum numbers of fields involved in neutrino masses.}
\label{tab:neutrinomasses}
\end{table}

The quantum numbers in Tab.~\ref{tab:neutrinomasses} lead to the off-diagonal entries in Eq.~(\ref{eq:Mnu}). In particular $(m_D^\prime)_{ij}$, defined as
\begin{align}
(m_D^\prime)_{iI}&=Y_R^{iI} \bar S^i \tilde H_R N_R^I+h.c.\nonumber\\
(m_D^\prime)_{i3}&=Y_R^{i3} \bar S^i \tilde H_R L_R^3+h.c.
\end{align}
is a diagonal matrix, while also the matrix $(m_D)_{ij}$ is diagonal as the bi-doublet $\mathcal H$ does not carry any lepton number.
Moreover we will introduce the non-diagonal Majorana mass matrix for singlets as $M_{ij}S_L^i S_L^j$ which will constitute a soft breakdown of the global symmetry $U(1)_{L_e}\otimes U(1)_{L_\mu}\otimes U(1)_{L_\tau}$, by the small $M$ mass matrix elements, leading to the neutrino mass matrix~\cite{Barr:2003nn}
\be
m_\nu=m_D \frac{1}{m_D^\prime}M \left(m_D \frac{1}{m_D^\prime}\right)^T
\ee
which should describe the neutrino masses and PMNS mixing angles~\cite{Patrignani:2016xqp}.

\section{Generating $R_{D^{(\ast)}}$}
\label{sec:RD}

Only fermion doublets localized on the IR brane, with both non-vanishing components, will interact with $W_R$. Then we can write the 4D charged current Lagrangian, Eq.~(\ref{eq:lagch}), in the mass eigenstate fermion basis as
\be
\mathcal L = \frac{g_R}{\sqrt{2}}\sum_{n=1}^\infty\left\{ \bar u_R (V_{u_R}^\dagger G^n\, V_{d_R}) \slashed{W}_R^n d_R+\bar\tau_R \slashed{W}_R^n G^n\nu_{\tau_R} \right\}
\label{eq:Wuu}
\ee
where the matrix form has been used. The coupling matrix $G^n$ can be approximated by
\be
G^n\equiv\diag\left(G^n_1,G^n_2,G^n_3
\right) 
\ee
where $G_{1,2}^n\ll G^n_3=f_{W_R}^n(y_1)$, and $f_{W_R}^n(y)$ is the normalized wave-function of the Kaluza-Klein modes of $W_R^n$ (see App.~\ref{sec:modes}). After integration of the KK modes we can write down the effective Lagrangian
\be
\mathcal L_{eff}=-\frac{4 G_F}{\sqrt{2}}V_{cb}C_\tau (\bar c_R\gamma^\mu b_R)(\bar\tau_R\gamma_\mu\nu_{\tau_R})
\ee
which has been normalized to the SM contribution, where the Wilson coefficient is given by
\be
C_\tau=\sum_n\left(\frac{g_R}{2}\,G_3^n\frac{v}{m_{n}}\right)^2\frac{\left(V_{u_R}^\dagger\right)_{23}}{V_{cb}}\simeq 1.45 \left(\frac{g_R}{2}\,G_3\frac{v}{m_{1}}\right)^2\frac{\left(V_{u_R}^\dagger\right)_{23}}{V_{cb}}
\label{eq:Ctau}
\ee
where $G_3\equiv G_3^1$ and $m_{1}$ are the coupling and mass of the first KK mode, and the pre-factor 1.45 takes into account the contribution of the whole tower. 

\begin{figure}[htb]
\centering
\includegraphics[width=10cm]{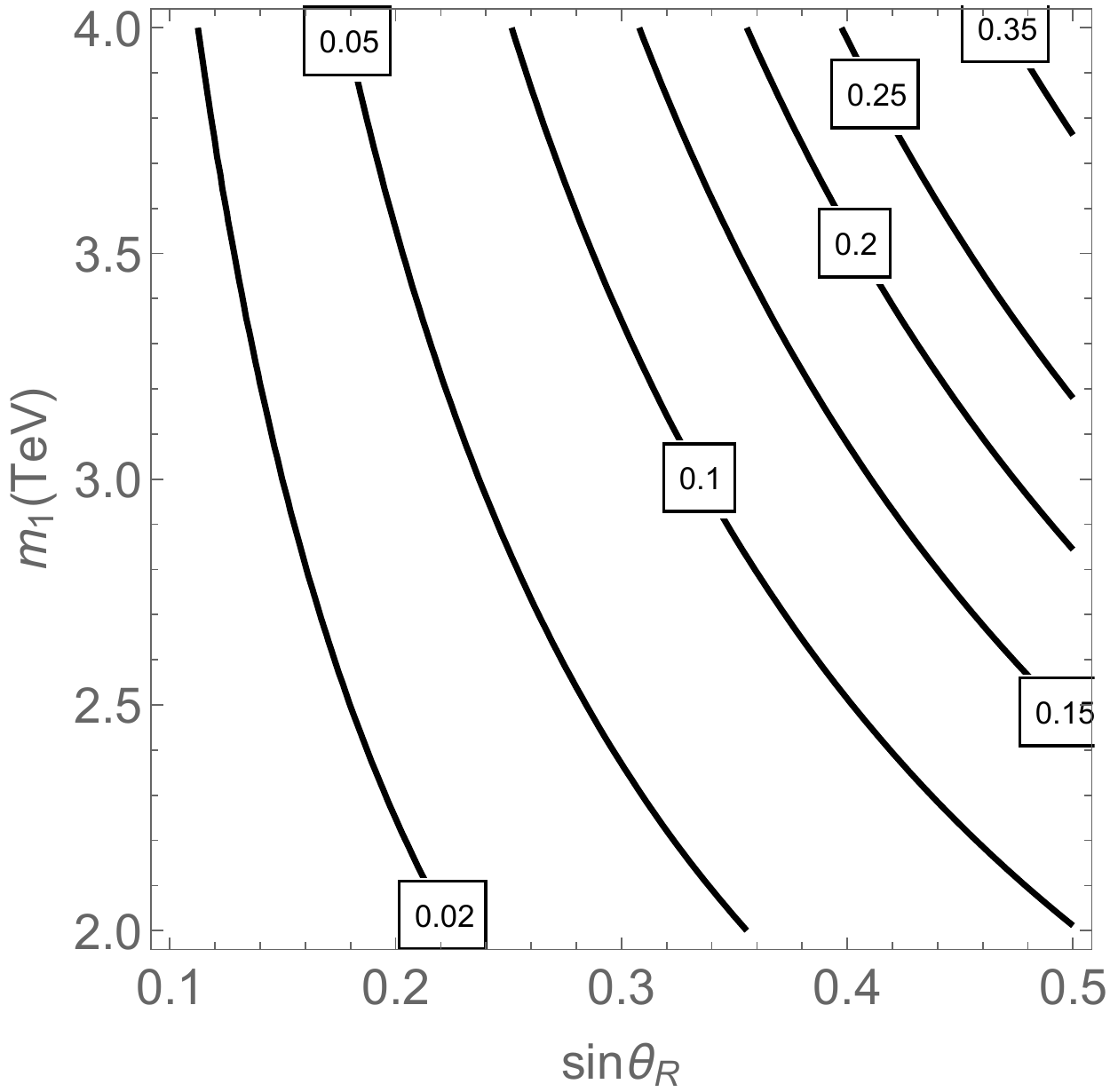} 
\caption{\it Contour lines of $(V_{u_R}^\dagger)_{23}$ in the plane $(\sin\theta_R,m_{1})$ as fixed from the best fit value to the experimental data for $R_D^{(\ast)}$.
}
\label{fig:VuR}
\end{figure}

The Wilson coefficient $C_\tau$ contributes to the process $b\to c \tau \bar \nu_{\tau}$ and thus to the ratio
\be
\frac{R_{D^{(\ast)}}}
{R_{D^{(\ast)}}^{SM}}=1+|C_\tau|^2
\ee
where
\be
R_{D}^{SM}=0.300\pm 0.011,\quad R_{D^{\ast}}^{SM}=0.254\pm 0.004
\ee
is the SM prediction~\cite{Aoki:2016frl,Amhis:2016xyh,Jaiswal:2017rve,Tran:2018kuv}, 
and the best fit value to experimental data is given by $C_\tau\simeq 0.46$~\cite{Greljo:2018ogz}~\footnote{In Ref.~\cite{Greljo:2018ogz} the best fit value $C_\tau\simeq 0.46$ is shown to be consistent with the experimental bound $\mathcal B(B_c\to\tau\bar\nu)<0.05$~\cite{Alonso:2016oyd}.}. Using this value there is a relation between the ratio $\left(V_{u_R}^\dagger\right)_{23}/V_{cb}$ and the mass $m_{1}$ given by
\be
m_{1}\simeq \frac{0.64}{\sin\theta_R}\, \left(\frac{\left(V_{u_R}^\dagger\right)_{23}}{V_{cb}}\right)^{1/2}  \textrm{ TeV}
\label{eq:WRmass}
 \ee
so that the element $\left(V_{u_R}^\dagger\right)_{23}$ as a function of $\sin\theta_R$ and the mass $m_{1}$ is given in Fig.~\ref{fig:VuR}.

In principle the anomaly in the branching ratio $\mathcal  B(B\to D^{(\ast)}\tau_R\bar\nu_R)$ might give rise to a large contribution to the branching fraction $\mathcal B(D_s\to \tau\bar\nu)\simeq 0.05$ from the process $\bar s c\to  \tau_R^+ \nu_R$, which is mediated by the KK modes $W_R^n$. However since $c_R$ and $s_R$ are in the bulk, and in different $SU(2)_R$ doublets, they  couple to $W_R^n$ only via mixing with the third generation quarks.  This implies that this contribution is further suppressed by a  factor 
$(V_{d_R})_{32}$ which, as we will see, is restricted to be small to satisfy the constraints on $\Delta m_{B_s}$.  Thus,  no significant contribution to the branching ratio $\mathcal B(D_s\to \tau\nu)$ is obtained. 

Similarly, in this model one would also expect an excess in the observable
\begin{equation}
R(J/\Psi) = \frac{B(B_c^+ \to J/\Psi\ \tau^+ \nu_\tau)}{B(B_c^+ \to J/\Psi\ \mu^+ \nu_\mu)}.
\end{equation}
The LHCb experiment has recently provided a result on this observable, showing an excess of the order of  2~$\sigma$ above the SM expected value,
$R(J/\Psi)_{\rm SM} \simeq 0.25$--$0.28$, Ref.~\cite{Aaij:2017tyk,Huang:2018nnq},
with large errors
\begin{equation}
R(J/\Psi) = 0.71 \pm 0.25.
\end{equation} 
Theoretical analyses of this observable~\cite{Issadykov:2018myx,Cohen:2018dgz} confirm this anomaly and 
show it to be governed by the same operator as the one governing $R_{D^{(*)}}$. In our particular model, we have 
\begin{equation}
\frac{R(J/\Psi)}{R(J/\Psi)_{\rm SM}} = 1 +  |C_\tau|^2\ .
\end{equation}
Given the value of $R(J/\Psi)_{\rm SM}$, the  measured value of this ratio is about  $2.6\pm 1$. Hence, the value of $C_\tau$ obtained above to explain $R_{D^{(*)}}$ can only slightly ameliorate this anomaly, and one should wait for more accurate experimental measurements of $R(J/\Psi)$ before further discussion of this issue.

\section{Constraints}
\label{sec:Constraints}
In this section we will examine the main constraints in processes which are related to $R_{D^{(\ast)}}$, and where the strong coupling of the third generation RH quarks and leptons to KK modes plays a significant role. To do that one has to compute the mixing between the electroweak gauge bosons $W^\pm_L$ and $Z_L$ and the KK modes using the effective Lagrangian.

We can easily compute the effective description of the Lagrangian, with mixing terms $W_LW_{L,R}^n$ and $Z_L Z_{L,R}^n$, generated by the 
vacuum expectation values of the bulk Higgs bi-doublets $\mathcal H$ and $\Sigma$  as well as  the Higgs doublet $H_R$ in the representation $(1,2)$, with VEV $\langle H_R\rangle=(v_R,0)^T$, and with $Q_X=-1/2$.
These are induced from the kinetic terms in the 5D Lagrangian as
\begin{align}
\mathcal L^{\rm GH}&= \tr\left|g_L W_L^aT_L^a \mathcal H-g_R \mathcal H W_R^a T_R^a
\right|^2+ \tr\left|g_L W_L^aT_L^a \Sigma-g_R \Sigma W_R^a T_R^a-g_X X\Sigma
\right|^2 \nonumber \\
&+\left|g_R H_R W_R^aT_R^a-\frac{1}{2}g_X XH_R \right|^2
\end{align}
where we are using the fact that $T_L^a$ acts on the bi-doublets rows and $T_R^a$ on the bi-doublets columns.

A straightforward calculation gives for the 4D quadratic Lagrangian for the gauge boson $n$-th KK modes 
\begin{align}
&\mathcal L_n^{\rm G}=g_L^2\,\frac{v^2}{4}  W_L W_L+ \frac{g_L^2}{\cos^2\theta_L}\frac{v^2}{8} Z_LZ_L \label{eq:quadratic}\\
&+
\frac{v^2}{4}  G_3^n r_h(\alpha)\bigg\{
g_L^2 (W_L W_L^n+h.c.)-\frac{2 v_1 v_2}{v^2}\,  g_L g_R(W_L W_R^n+h.c.)\bigg\}\nonumber\\
&+\frac{v^2}{4}G_3^n r_h(\alpha)\bigg\{
\frac{g_L^2}{\cos^2\theta_L} Z_L Z_L^n+
\frac{g_Lg_R}{\cos\theta_L\cos\theta_R}\left[2\,\sin^2\theta_\Sigma-  \cos^2\theta_R\right]  Z_L Z_R^n\bigg\}
\nonumber
\end{align}
where $v^2=v_1^2+v_2^2+v_\Sigma^2$, the first two terms provide the $W_L$ and $Z_L$-masses, and we have introduced the function $r_h(\alpha)$ which depends on the localization in the bulk of the $h$ Higgs direction acquiring a vacuum expectation value. In fact for a Higgs localized in the IR brane, $\alpha\to\infty$, one gets $r_h\simeq 1$, while for a Higgs localized towards the UV brane $\alpha\leq 1$ one gets $r_h\simeq 0$. For $\alpha = 2$ the Higgs is sufficiently localized towards the IR brane to solve the hierarchy problem, and  we shall use this value in the rest of this article, leading to a  factor  $r_h\simeq 0.68$.
%

Another important effect for analyzing the relevant constraints, in the presence of composite, and partly composite, fermions $f$, is that in our model the effective operators
\be
\mathcal O_{f t_R}=(\bar f\gamma^\mu f)(\bar t_R\gamma_\mu t_R)
\label{eq:OfRtR}
\ee
are induced, with Wilson coefficients given by
\be
C_{f t_R}= -\sum_n \left(\frac{G_3^n}{m_n}\right)^2 r_f(c_f)\left[\frac{g_L^2}{\cos^2\theta_L}\left(
g_{Aff}g_{At_R t_R}+g_{Z_Lff}g_{Z_Lt_Rt_R}\right)+\frac{g_R^2}{\cos^2\theta_R}g_{Z_Rff}g_{Z_Rt_Rt_R}
\right].
\label{eq:CfRtR}
\ee
In the above,  we have introduced the function $r_f(c_f)$ as $$r_f(c_f)\equiv G_f^n(c_f)/G_3^n\ ,$$ where 
$G_f^n(c_f)$ is the overlapping integral of fermion zero mode profiles, for the given value of the $c_f$ parameter, and the 
gauge boson KK mode profile. In particular, for IR localized fermions, which could be considered as the limiting case where $c_f\to -\infty$, it turns out that $\lim_{c_f\to -\infty} r_f(c_f)=1$.
The Wilson coefficients trigger a one-loop modification of the $Z_L\bar ff$ couplings,
through a top-quark loop diagram followed by emission of the $Z_L$ gauge boson~\cite{Feruglio:2017rjo},
 which in turn induces the modification of the corresponding $Z_L\bar ff$ coupling. In particular, for the relevant cases we will analyze here $f=\tau_R,b_R,b_L,\mu_L$ are the composite ($b_R,\tau_R$), or partly composite ($b_L,\mu_L$), fermions.

\subsection{The coupling $Z\tau_R\tau_R$}
\label{sec:Ztautau}
As the $\tau_R$ lepton is localized on the IR brane, and it couples strongly to the KK modes, the main constraint will be the modification of the coupling $Z_L\tau_R\tau_R$, defined as
\be
\mathcal L_{Z\tau_R\tau_R}=\frac{g_L}{\cos\theta_L}\bar \tau_R\slashed{Z}_L(g_{Z_L\tau_R\tau_R}+\delta g_{Z_L\tau_R\tau_R})\tau_R ,
\label{eq:ZLtauRtauR}
\ee
where the term $\delta g_{Z_L\tau_R\tau_R}$ is constrained by the global fit to the experimental data of Ref.~\cite{Falkowski:2017pss} as
\be
\delta g_{Z_L\tau_R\tau_R}=(0.42\pm 0.62)\times 10^{-3}.
\ee

The term $\delta g_{Z_L\tau_R\tau_R}$ in Eq.~(\ref{eq:ZLtauRtauR}) is generated at the tree level by the mixing $Z_{L,R}^n Z_L$ induced by the Higgs vacuum expectation value,
 \begin{figure}[htb]
\centering
\includegraphics[width=10cm]{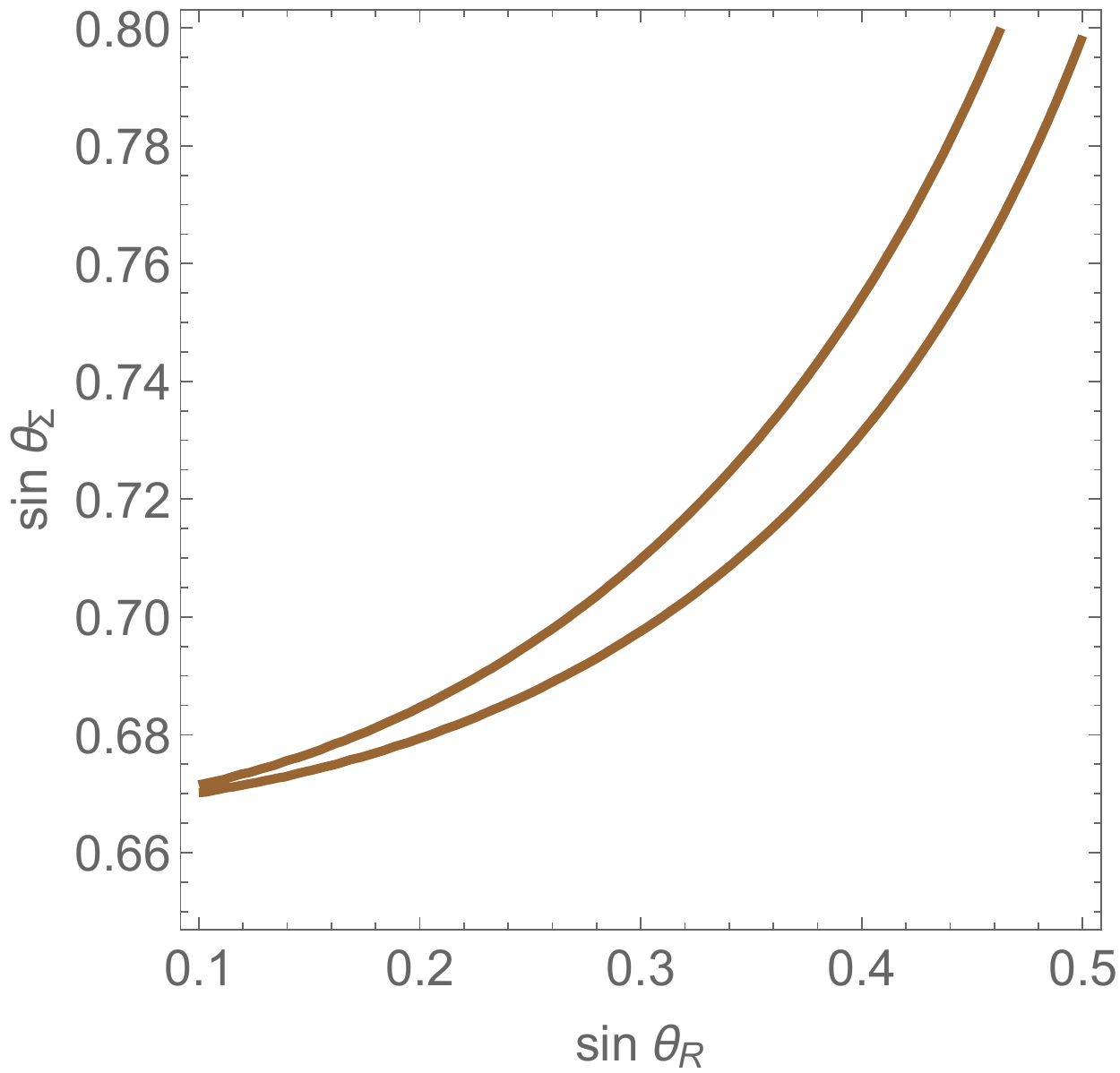} 
\caption{\it The region between the (brown) solid lines is allowed by the best fit to $\delta g_{Z_L\tau_R\tau_R}$ for $m_{1}=3$ TeV.
}
\label{fig:deltagZtauR}
\end{figure} 
and through radiative corrections using the effective operator 
\be
 \mathcal O_{\tau_R t_R}=(\bar\tau_R\gamma^\mu \tau_R)(\bar t_R\gamma_\mu t_R)
\label{eq:optauRtR}
\ee
with Wilson coefficient given by Eq.~(\ref{eq:CfRtR}).
Using now the mixing terms from Eq.~(\ref{eq:quadratic}) and the couplings from Eqs.~(\ref{eq:gZR}) and (\ref{eq:gZL}) we can write
\begin{align}
&\delta g_{Z_L\tau_R\tau_R}= \sum_n \left( \frac{g_R v G_3^n}{2 m_n} \right)^2\Bigg\{ \frac{r_h(\alpha)}{\cos^2\theta_R}\left[\sin^2\theta_\Sigma\left(1-2\sin^2\theta_R \right)-\frac{1}{2}\cos^2\theta_R
\right]\\
&+ \frac{3h_t^2}{4\pi^2}\log\frac{m_1}{m_t}
\left[ \frac{2}{3}\sin^2\theta_R+\frac{1}{\cos^2\theta_R}\left(\frac{1}{2}-\sin^2\theta_R\right)\left(\frac{1}{2}-\frac{2}{3}\sin^2\theta_R\right) \right]\Bigg\}, \nonumber
\end{align}
%
where the first line comes from the contribution of the KK gauge bosons through mixing effects and the second line is the radiative contribution from the top quark loop~\footnote{We have done the calculation using DimReg and the $\overline{MS}$ renormalization scheme.} induced by the operator (\ref{eq:optauRtR}). The
coupling $h_t$ is the SM top-Yukawa coupling, defined by
\begin{equation}
m_t \equiv h_t\, v/\sqrt{2}\,,
\end{equation}
which is therefore related to the $Y_{33}^u$ coupling defined in Eq.~(\ref{eq:Yu}) by
\begin{equation}
h_t = \cos\theta_\Sigma  Y^u_{33} \,.
\end{equation}

In order to determine the KK-mode contribution we use
the condition (\ref{eq:Ctau}) on $R_{D^{(\ast)}}$ and get the allowed region in the plane $(\sin\theta_\Sigma,\sin\theta_R)$ shown in Fig.~\ref{fig:deltagZtauR}, where we are assuming $m_1=3$ TeV. 
Fig.~\ref{fig:deltagZtauR} shows that the constraint on $\delta g_{Z_L\tau_R\tau_R}$ puts a lower bound on $\sin\theta_\Sigma$, which is given by 
\be
 \sin\theta_\Sigma= \frac{v_\Sigma}{v}\gtrsim 0.67\,,
\ee
and in particular excludes the value $\sin\theta_\Sigma=0$, i.e.~it requires the introduction of the Higgs bi-doublet $\Sigma$.

\subsection{Oblique observables}
\label{sec:T}
In these theories the $T$-parameter, defined as,
\be
\alpha_{EM}(m_Z)T=\left[\frac{\Pi_{WW}(0)}{m_W^2}-\frac{\Pi_{ZZ}(0)}{m_Z^2}\right],
\ee
is protected by the custodial symmetry in the bulk only in the case when $\tan\beta=1$ and $\sin\theta_\Sigma=0$. 

In general, there may be relevant contributions to the precision electroweak observables induced by the mixing of the gauge boson zero modes with the KK modes, as given by Eq.~(\ref{eq:quadratic}), as well as loop corrections induced by top loop corrections. In fact in a similar way as the operator
(\ref{eq:OfRtR}) is generated by exchange of $(A^n,Z_L^n,Z_R^n)$ KK modes, the operator
\be
(H^\dagger iD_\mu H)(\bar t_R \gamma^\mu t_R)
\label{eq:operador}
\ee
is generated by the mixing of $Z_L$ with KK modes in (\ref{eq:quadratic}) followed by the exchange of $(Z_L^n,Z_R^n)$ KK modes coupled to the top quark. The radiative correction to the $T$ parameter is obtained after closing the top-loop, and by emission of a $Z_L$-gauge boson from it. 

There are also loop contributions involving fermionic KK modes, but
in a scenario in which the right handed third generation fermions are localized on the infrared brane, they  strongly depend on the localization 
of the left handed third generation quarks (see, for example, Refs.~\cite{Carena:2004zn,Carena:2006bn,Carena:2007ua}).  In particular, these loop corrections are strongly suppressed when the left-handed third generation quarks are 
localized close to the IR brane, or in the presence of sizable quark brane kinetic terms.  Moreover, unlike the mixing between gauge KK $n$-modes and gauge zero modes, which is enhanced for IR brane localized fermions by $\sim |G_3^n|=\sqrt{2k y_1}$, the mixing between fermion KK $n$-modes and fermion zero modes is $\sim G_3^n/\sqrt{ky_1}$, so that the loop corrections to the $T$ parameter are not volume-enhanced, while they are suppressed by the mass of the heavy fermions and by loop factors. 
Hence, in this work, we shall concentrate 
on the relevant corrections to flavor physics observables induced by the gauge boson mixing, and the inter-generational mixing of the right-handed
quarks, as well as by the top loop corrections we have just described from the operator (\ref{eq:operador}). These corrections to the precision electroweak observables  are well defined within our framework, and are strongly correlated with our proposed solution to $R_{D^{(*)}}$.

 \begin{figure}[htb]
\centering
\includegraphics[width=10cm]{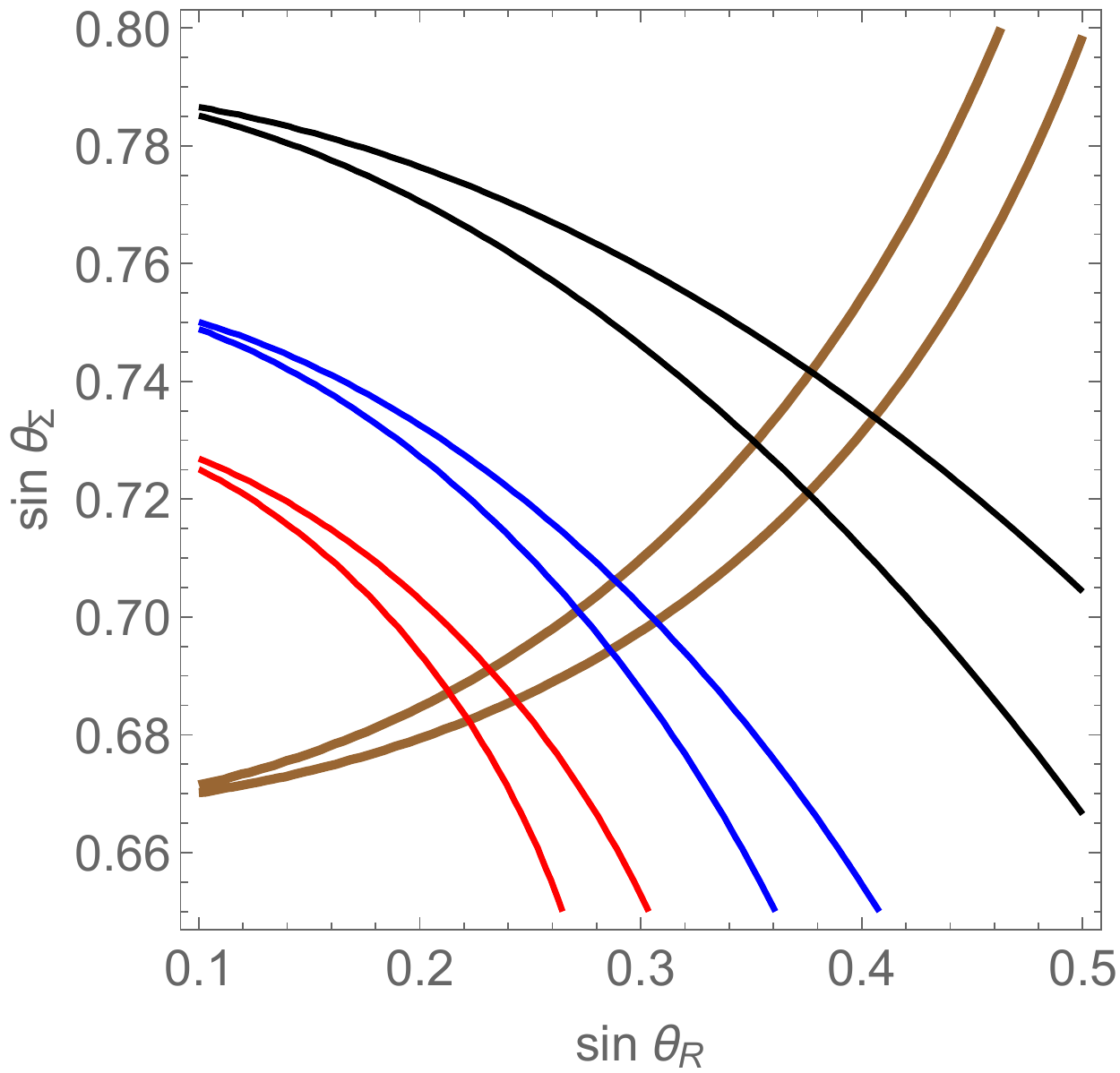} 
\caption{\it The region between the solid lines is allowed by $\delta g_{Z_L\tau_R\tau_R}$ (brown lines, as obtained in Fig.~\ref{fig:deltagZtauR}) and by 
the 95 \% C.L. bound on the $T$  parameter, for  $m_{1}=3$ TeV and $\tan\beta=1$ (black lines), $\tan\beta=3$ (blue lines) and $\tan\beta=5$ (red lines). 
}
\label{fig:alphaT}
\end{figure} 
We can easily compute the contributions to the $T$-parameter induced by the mixing of the zero mode gauge bosons with the KK modes by using the effective description of the Lagrangian, with mixing terms $W_LW_{L,R}^n$ and $Z_L Z_{L,R}^n$, from Eq.~(\ref{eq:quadratic}), and at one-loop from the effective operator (\ref{eq:operador}).  
Working to lowest order, $\mathcal O(v^4)$, in Higgs insertions, we obtain the result
\begin{align}
\alpha T&=
r_h(\alpha) \sum_n\left( \frac{g_R\,v\, G_3^n}{2 m_{n}} \right)^2\,\Bigg\{ r_h(\alpha)\bigg[\cos^2 2\beta -4\sin^2\theta_\Sigma\left(1-\frac{\sin^2\theta_\Sigma }{\cos^2\theta_R}  \right)\nonumber\\
&+\sin^2 2\beta \sin^2\theta_\Sigma\left(2-\sin^2\theta_\Sigma\right)
\bigg]\label{eq:Texpression}\\
&-\frac{3h_t^2}{2\pi^2}\log\frac{m_1}{m_t}\bigg[-\frac{2}{3}\sin^2\theta_R+\frac{1}{\cos^2\theta_R}
\left(2\sin^2\theta_\Sigma-\cos^2\theta_R  \right)\left(\frac{1}{2}-\frac{2}{3}\sin^2\theta_R  \right)
\bigg]\Bigg\}\nonumber
\end{align}
where the first two lines is the tree-level result and the third line the radiative correction induced at one-loop by the mixing between the tree-level (\ref{eq:quadratic}) and one-loop (\ref{eq:operador}) operators.

 Using now the expression fitting the value of $R_{D^{(\ast)}}$, we can obtain the allowed regions for the $T$ parameter in the $(\sin\theta_\Sigma, \sin\theta_R)$ plane, fixing the
 values of $m_1$ and $\tan\beta$.  In Fig.~\ref{fig:alphaT}, in addition to the $\delta g_{Z_L\tau_R\tau_R}$ bounds from Fig.~\ref{fig:deltagZtauR}, we show the regions allowed by
 the $T$ parameter experimental bounds at the 95\% confidence level~\cite{Patrignani:2016xqp}
%
%
\be
T=0.07\pm 0.12,
\label{eq:Texp}
\ee
for $m_{1}=3$ TeV and several values of $\tan\beta=1,\, 3,\, 5$. The value $T=0$ is a middle line inside every band. 
In order to reduce the value of the Yukawa coupling $Y^u_{33}$, Eq.~(\ref{eq:Yu}), we  should consider values of $\tan\beta > 1$.
The intersection of the $\delta g_{Z_L\tau_R\tau_R}$ allowed band with the  $T$ parameter allowed band for $\tan\beta = 1$
(solid brown and black lines, respectively), define the 
upper bounds on $\sin\theta_R$ and $\sin\theta_\Sigma$ in the
regime we are considering as,
\be
\sin\theta_R\lesssim 0.4\,,\quad \sin\theta_\Sigma\lesssim 0.75\ .
\ee
%

%
 \begin{figure}[htb]
\centering
\includegraphics[width=7.5cm]{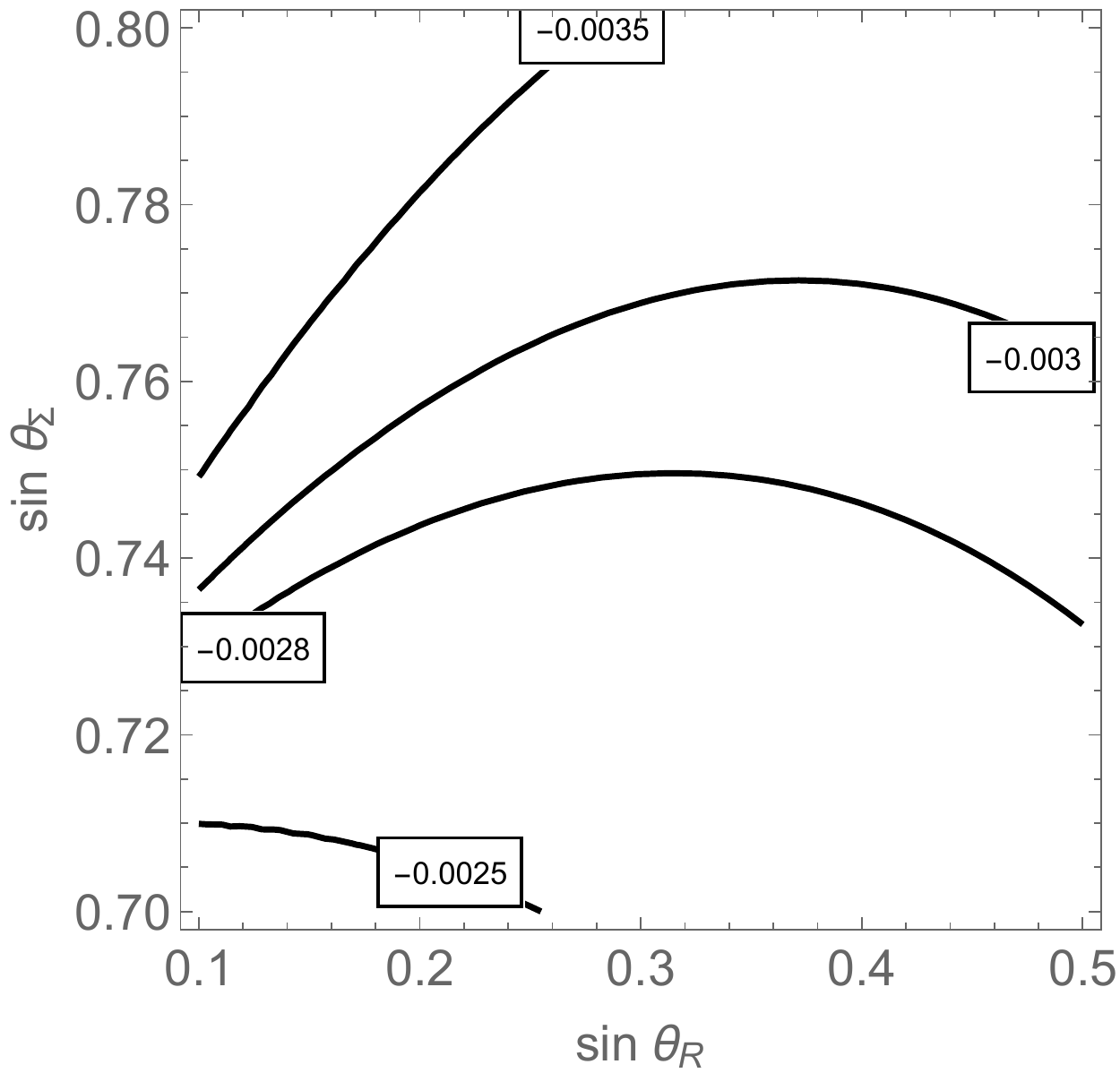} 
\includegraphics[width=7.5cm]{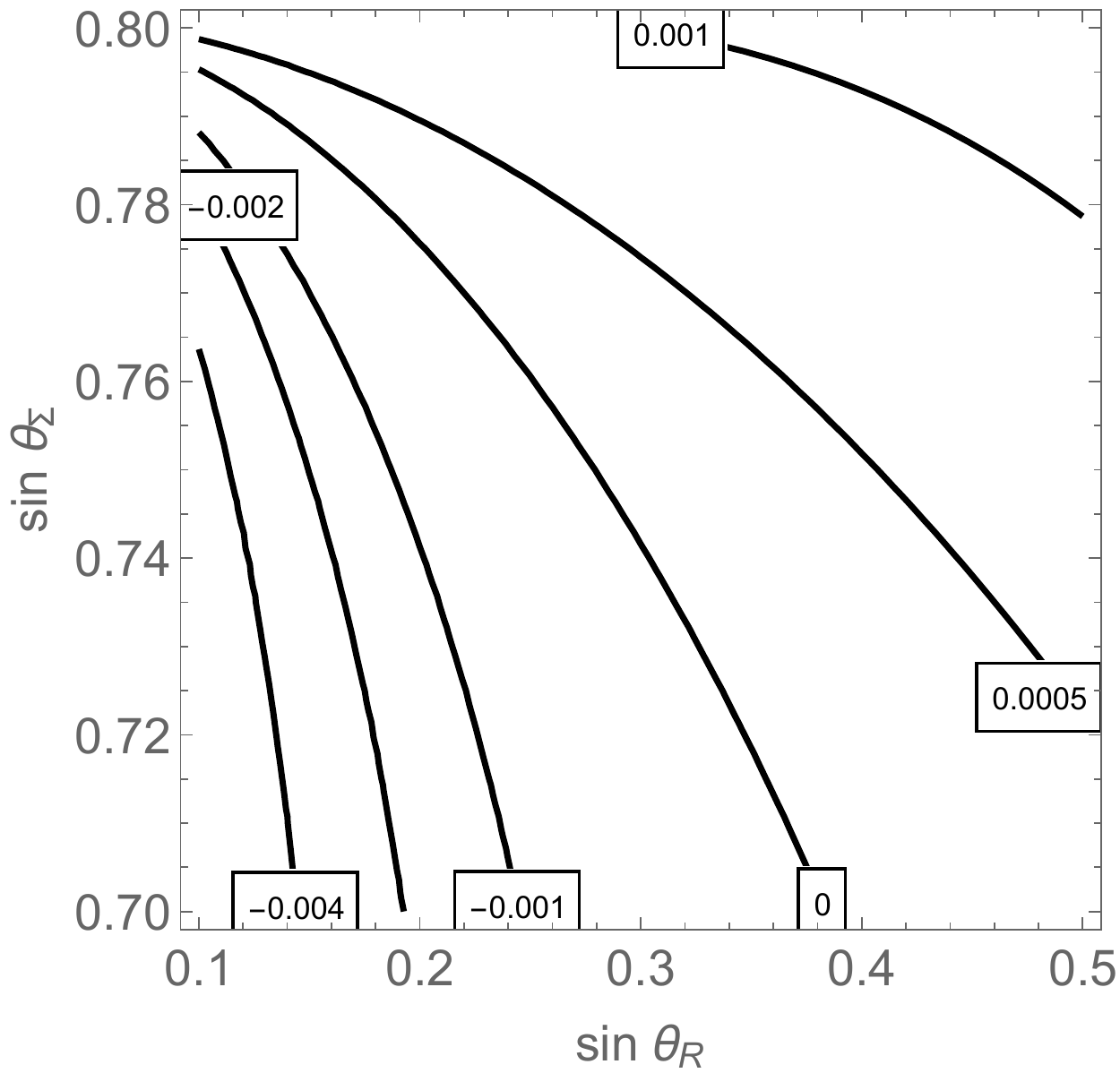}
\caption{\it Contour lines of $S$ (left panel) and $U$ (right panel) for the case $m_{1}=3$ TeV and $\tan\beta=2$. 
}
\label{fig:S}
\end{figure} 
As for the $S$ and $U$ parameters, they are defined in our theory as
\begin{align}
\alpha_{EM}(m_Z) S&=4\sin^2\theta_L \cos^2\theta_L \Pi^\prime_{ZZ}(0)\\
\alpha_{EM}(m_Z) (S+U)&=4\sin^2\theta_L  \Pi^\prime_{WW}(0)
\end{align}
which, using the effective description of Eq.~(\ref{eq:quadratic}), can be cast as
\begin{align}
\alpha_{EM}(m_Z) S &=-\frac{2 r_h^2(\alpha)}{ky_1}\sum_n\left(\frac{v g_R G_3^n}{2 m_n} \right)^4
\sin^2\theta_R \cos^2\theta_L \nonumber \\
&\quad \times \left[\frac{\sin^2\theta_R}{\sin^2\theta_L}+\frac{\left( 2\sin^2\theta_\Sigma -\cos^2\theta_R\right)^2}{\cos^2\theta_R}
\right]
\label{eq:S}
\end{align}
and
\begin{align}
\alpha_{EM}(m_Z) (S+U) &= -\frac{2 r_h^2(\alpha)}{ky_1}\sum_n\left(\frac{v g_R G_3^n}{2 m_n} \right)^4
\sin^2\theta_R \cos^4\theta_L  \nonumber \\
&\quad \times \left[\frac{\sin^2\theta_R}{\sin^2\theta_L}+\frac{\sin^2 2\beta}{\cos^2\theta_L}\cos^4\theta_\Sigma 
\right] \,,
\label{eq:S}
\end{align}
where, as their tree level values are so small, we are neglecting its crossing with the radiative corrections induced by the operator (\ref{eq:operador}). 

After applying the constraint from the $R_{D^{(\ast)}}$ anomaly, 
fixing the value of the KK mass, $m_{1}=3$ TeV, and $\tan\beta=2$, the $S$ and $U$ contours are depicted in
Fig.~\ref{fig:S}. It follows from this figure 
that the predicted values are consistent with the experimental constraint~\cite{Patrignani:2016xqp}
\be
S=0.02\pm 0.10,\quad U=0.00\pm 0.09
\ee
in all the parameter region. Similar small values of $S$ and $U$ are obtained for other values of $\tan\beta$.

\subsection{Flavor observables}
\label{sec:flavor}
New physics contribution to $\Delta F=2$ observables appears mainly from exchange of KK gluons. The leading flavor violating couplings of the KK gluons $\mathcal G_\mu^n$ involving RH down and up quarks is given by
\be
\mathcal L_s=g_s (V^\dagger_{u_R})_{i3} (V_{u_R})_{3j}\bar u^i_R \slashed{\mathcal G}^n G^n_3 u_R^j +g_s (V^\dagger_{d_R})_{i3} (V_{d_R})_{3j}\bar d^i_R \slashed{\mathcal G}^n G^n_3 d_R^j.
\ee
After integrating out the gluon KK modes we obtain a set of $\Delta F=2$ dimension six operators. 
In particular, the most constrained operators are those given by
\be
\mathcal L_{eff}=C_{sd}(\bar s_R\gamma^\mu d_R)^2+C_{cu}(\bar c_R\gamma^\mu u_R)^2+
C_{bd}(\bar b_R\gamma^\mu d_R)^2+C_{bs}(\bar b_R\gamma^\mu s_R)^2
\ee
where the Wilson coefficients are given by
\begin{align}
C_{sd}&=\frac{g_s^2}{6}\left[ (V^\dagger_{d_R})_{23} (V_{d_R})_{31} \right]^2 \sum_n \left(
\frac{G_3^n}{m_n}\right)^2\label{eq:Csd}\\
C_{cu}&=\frac{g_s^2}{6}\left[ (V^\dagger_{u_R})_{23} (V_{u_R})_{31} \right]^2 \sum_n \left(
\frac{G_3^n}{m_n}\right)^2\label{eq:Ccu}\\
C_{bd}&=\frac{g_s^2}{6}\left[ 
 (V_{d_R})_{31} \right]^2 \sum_n \left(
\frac{G_3^n}{m_n}\right)^2\label{eq:Cbd}\\
C_{bs}&=\frac{g_s^2}{6}\left[ 
(V_{d_R}^\dagger)_{23} \right]^2 \sum_n \left(
\frac{G_3^n}{m_n}\right)^2,
\label{eq:Cbs}  
\end{align}
where $(V_{u_R}^\dagger)_{23}$ is constrained by $R_{D^{(\ast)}}$, see Fig.~\ref{fig:VuR}. 
If, for simplicity, we assume real matrices $V_{u_R}$ and $V_{d_R}$ (no CP violation in the right-handed sector) the Wilson coefficients $C_{sd}$, $C_{cu}$, $C_{bd}$ and $C_{bs}$ are constrained from $\Delta m_K$, $\Delta m_D$, $\Delta m_{B_d}$ and $\Delta m_{B_s}$, respectively, as~\cite{Isidori:2015oea,Ligeti:2015kwa}
\begin{align}
C_{sd}&<9\times 10^{-7} \textrm{ TeV}^{-2},\label{eq:boundCsd}\\
C_{cu}&<5.6\times 10^{-7} \textrm{ TeV}^{-2}, \label{eq:boundCcu} \\
C_{bd}&<2.3\times 10^{-6} \textrm{ TeV}^{-2},\label{eq:boundCbd}\\
C_{bs}&<5\times 10^{-5} \textrm{ TeV}^{-2} \label{eq:boundCbs}. 
\end{align}

\begin{figure}[htb]
\centering
\includegraphics[width=10cm]{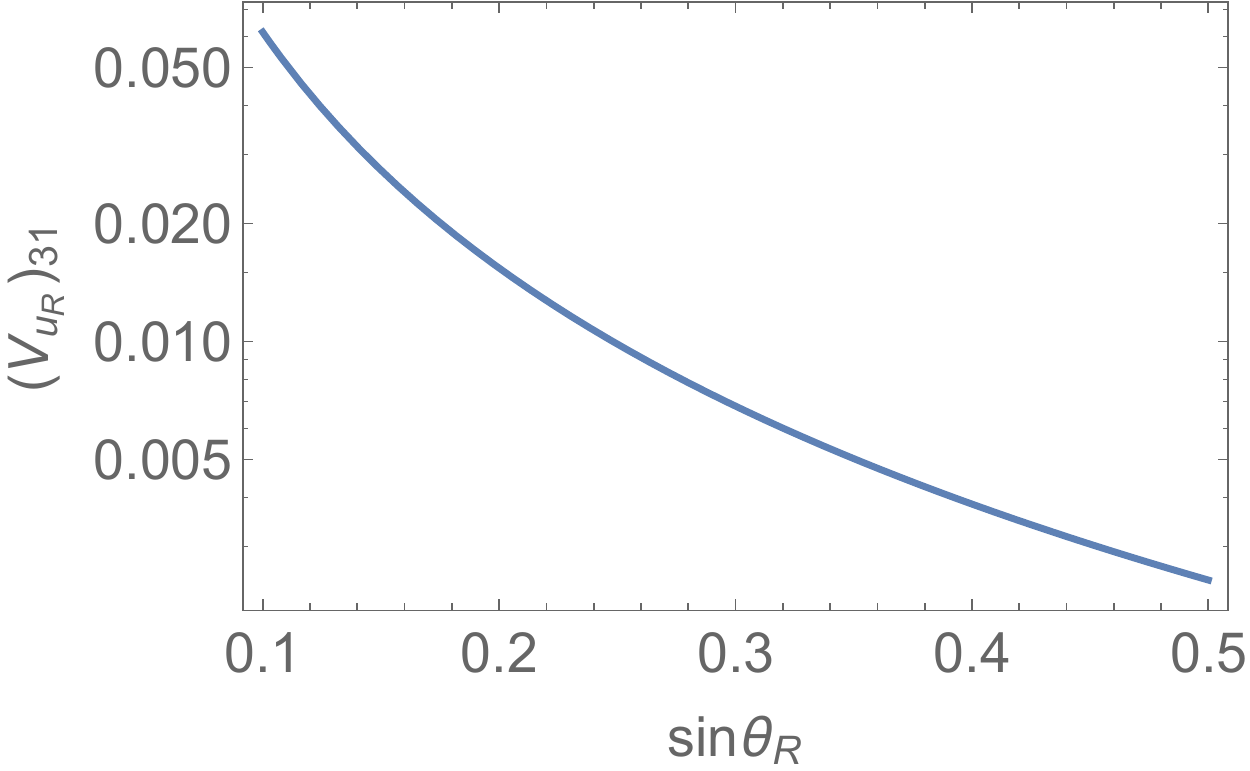} 
\caption{\it Upper bound on $(V_{u_R})_{31}$ as a function of $\sin\theta_R$ from condition (\ref{eq:boundCcu}) for $m_1$=3~TeV, using the value of $(V^\dagger_{u_R})_{23}$ provided by $R_{D^{(\ast)}}$.
}
\label{fig:VuR31}
\end{figure}

Operators involving third generation quarks, although providing weaker bounds on the Wilson coefficients, are very constraining as they contain the element $(V^\dagger_{d_R})_{33}\simeq 1$. In particular the bounds on $C_{bd}$ and $C_{bs}$, Eqs.~(\ref{eq:boundCbd}) and (\ref{eq:boundCbs}), provide bounds on $(V_{d_R})_{31}$ and $(V^\dagger_{d_R})_{23}$, respectively, as
\be
|(V^\dagger_{d_R})_{13}|\lesssim 1.1\times 10^{-3}\left(\frac{m_{1}}{3\textrm{ TeV}} \right),\quad
|(V^\dagger_{d_R})_{23}|\lesssim 5.2\times 10^{-3}\left(\frac{m_{1}}{3\textrm{ TeV}} \right) .
\label{eq:Cbdbs}
\ee
Using now the bounds in Eq.~(\ref{eq:Cbdbs}) we can bound the element $C_{sd}$ as 
\be
C_{sd}<4.5\times 10^{-11}\left(\frac{m_{1}}{3\textrm{ TeV}}\right)^2\, \textrm{ TeV}^{-2} ,
\ee
which is a stronger bound than Eq.~(\ref{eq:boundCsd}). 
Moreover, from the definition of $C_{cu}$ in Eq.~(\ref{eq:Ccu}) and the corresponding bound (\ref{eq:boundCcu}), we can fix an upper bound on the element $(V_{u_R})_{31}$ using the value of $(V^\dagger_{u_R})_{23}$ provided by $R_{D^{(\ast)}}$. The result is plotted in Fig.~\ref{fig:VuR31}
as a function of $\sin\theta_R$.

\subsection{Lepton flavor universality tests}
There are two processes where lepton flavor universality has been tested to hold with a high accuracy. The first one is the ratio
\be
R_{D^{(\ast)}}^{\mu/e}=\frac{\mathcal B(B\to D^{(\ast)}\mu \bar\nu_\mu)}
{\mathcal B(B\to D^{(\ast)}e \bar\nu_e)}
\ee
which is constrained by experimental data to be $R_{D^{(\ast)}}^{\mu/e}\lesssim 1.02$~\cite{Pich:2013lsa}. In our model, the process
$\Gamma (b\to c W_R^\ast\to c \ell\bar\nu_\ell)=0$, for $\ell=(\mu,e)$, since only the third generation leptons couple to $W_R$. Hence, it follows that $R_{D^{(\ast)}}^{\mu/e}=R_{D^{(\ast)}}^{\mu/e}|_{SM}\simeq 1$ and the experimental bound is satisfied .

The second process is 
\be
R_\mu^{\tau/\ell}=\frac{\mathcal B(\tau\to \ell \nu_\tau\bar\nu_\ell)/\mathcal B(\tau\to \ell \nu_\tau\bar\nu_\ell)_{SM}}{\mathcal B(\mu\to e \nu_\mu\bar\nu_e)/\mathcal B(\mu\to e \nu_\mu\bar\nu_e)_{SM}},\quad (\ell=\mu,e)
\ee
which is constrained by experimental data to be $R_\mu^{\tau/\mu}=1.0022\pm 0.0030$ and 
$R_\mu^{\tau/e}=1.0060\pm 0.0030$. It turns out that the contribution to these processes from $W_R$, $\mathcal B(\tau\to \nu_\tau W_R^\ast\to \nu_\tau \ell \bar \nu_\ell)$ and similarly
$\mathcal B(\mu\to \nu_\mu W_R^\ast\to \nu_\mu \ell \bar \nu_\ell)$ is negligible for the same reason as before, and hence
the deviation of $R_\tau^{\tau/\ell}$ with respect to the SM values is also negligible, in good agreement with these measurements.

\subsection{LHC bounds}
The first neutral KK resonance $X^1$ ($X=Z_L,\, Z_R,\, A$) can be produced on-shell at LHC in Drell-Yan processes $\sigma(b\bar b\to X^1)$, followed by decays $X^1\to f\bar f$ where $f=\tau_R,b_R,t_R$. The production cross-section times branching ratio can be written as
\begin{align}
&\sum_X\sigma(pp\to X^1)\times \mathcal B(X^1\to f\bar f)=\frac{1}{9}g_R^2\, 2k y_1 f(m_{1})\Big[\sin^2\theta_R \sin^2\theta_L \mathcal B(Z_L^1\to f\bar f )
\nonumber\\
&
+\frac{1}{\cos^2\theta_R}(3/2-\sin^2\theta_R)^2 \mathcal B(Z_R^1\to f\bar f)
+\sin^2\theta_R\cos^2\theta_L \mathcal B(A^1\to f \bar f))
\Big]
\end{align}
where $f(m_{1})$ is the production cross-section for unit coupling obtained by  \texttt{MadGraph
  v5}~\cite{Alwall:2014hca}. 
\begin{figure}[htb!]
\centering
\includegraphics[width=10cm]{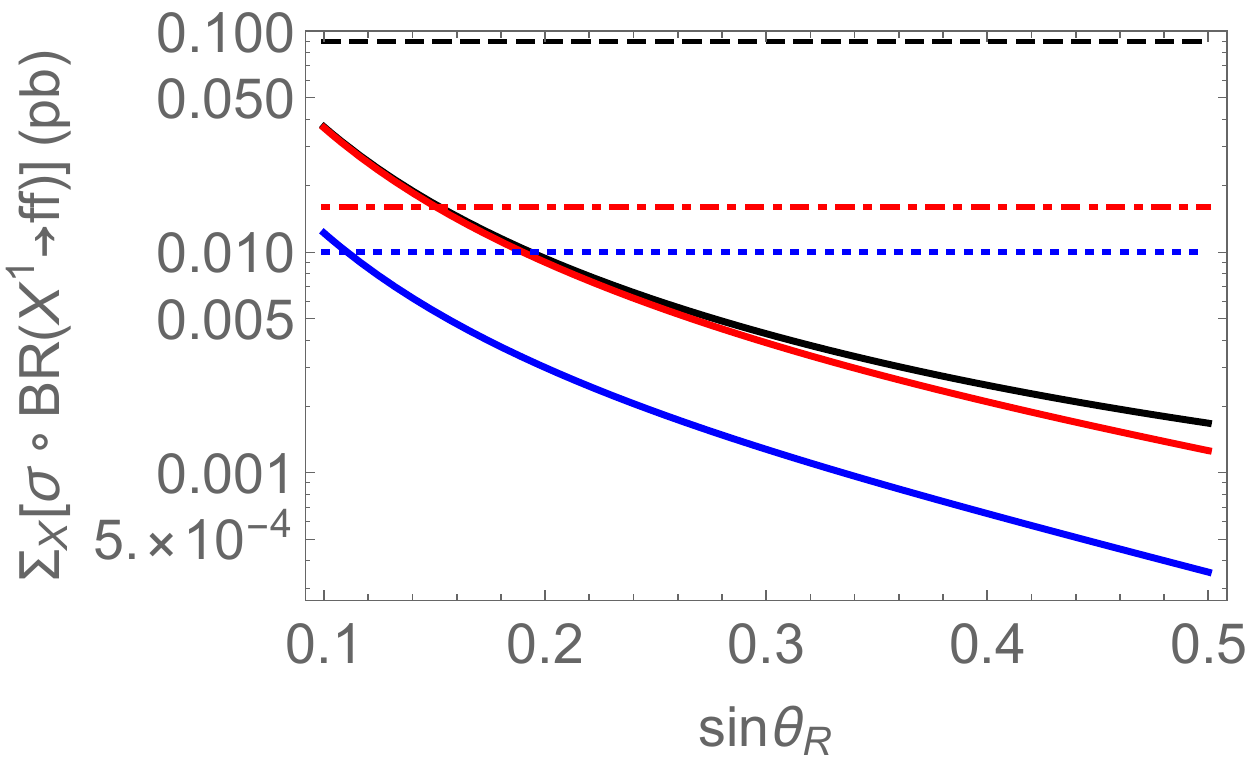} \hspace{0.5cm}
\caption{\it Plot of $\sum_X\sigma\times \mathcal B(X^1\to \bar ff)$ as a function of $\sin\theta_R$, for $m_{1}=3$ TeV and $f=b_R$ (upper black solid line), $f=t_R$ (middle red solid line) and $f=\tau_R$ (lower blue solid line). Horizontal lines correspond to the 95\% CL experimental upper bounds from the ATLAS experiment for $f=b_R$ (upper dashed line), for $f=t_R$ (middle dot-dashed red line) and $f=\tau_R$ (lower dotted blue line). 
}
\label{fig:sigmaff}
\end{figure} 

Our model prediction for $\sum_X\sigma(pp\to X^1) \times \mathcal B(X^1\to \bar f f)$
is given by the upper, middle and lower solid lines of Fig.~\ref{fig:sigmaff} 
for $f=b_R,\, t_R,\, \tau_R$, respectively.
We compare them with the experimental 
95\% CL upper bounds from the corresponding processes, which are given by the dot-dashed (red), dashed (black) and dotted (blue) horizontal lines from the ATLAS experiment on $\sigma\times \mathcal B(Z^\prime\to \bar tt)$~\cite{Aaboud:2018mjh}, $\sigma\times\mathcal B(Z^\prime\to \bar bb)$~\cite{Aaboud:2018tqo} and $\sigma\times\mathcal B(Z^\prime\to \tau\tau)$~\cite{Aaboud:2017sjh} for $m_{Z^\prime}=3$ TeV, respectively.
 As can be seen from Fig.~\ref{fig:sigmaff} only the process $\sigma\times\mathcal B(Z^\prime\to \bar tt)$ puts a significant bound on our model, of $\sin\theta_R\gtrsim 0.15$ for $m_{1}=3$ TeV, as we are assuming. These results, when extrapolated to masses of order 3~TeV, are consistent with those of the collider analysis presented 
 in Ref.~\cite{Greljo:2015mma}.

\begin{figure}[htb!]
\centering
\includegraphics[width=10cm]{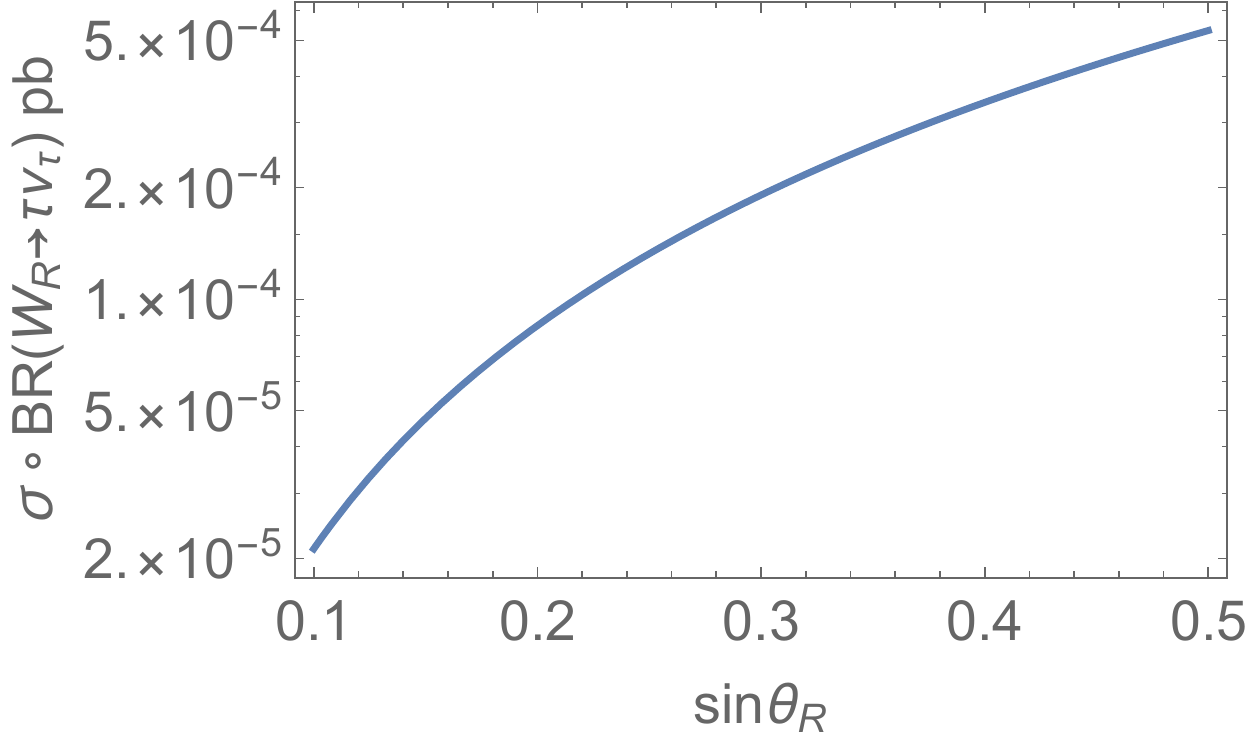} \hspace{0.5cm}
\caption{\it Plot of $\sigma (pp\to W_R^1)\times \mathcal B(W_R^1\to \tau_R \nu_{\tau_R})$ as a function of $\sin\theta_R$, for $m_{1}=3$ TeV
and the values of $(V_{u_R}^\dagger)_{23}$ required for the solution to the $R_D^{(*)}$ anomaly. 
}
\label{fig:sigmatauneutrino}
\end{figure}

In a similar way the first charged KK resonance $W_R^1$ can be produced on-shell at the LHC in the process $\sigma(b\bar c\to W_R^1)$, followed by the decays $W_R\to \tau_R\nu_{\tau_R},  t_R \bar b_R$, that assuming that there are no exotic fermions localized in the IR brane,
yield branching ratios around 1/4 and 3/4, respectively. In our model the production cross sections times branching-ratio is
\be
\sigma (pp\to W_R^1)\times \mathcal B(W_R^1\to \tau_R \nu_{\tau_R})\simeq \frac{g_R^2}{8} G_3^2 (V_{u_R}^\dagger)_{23}^2\, g(m_1)
\ee
where $g(m_{1})$ is the production cross-section for unit coupling obtained by  \texttt{MadGraph
  v5}~\cite{Alwall:2014hca}~\footnote{We thank Xiaoping Wang for help in the computation of these cross sections.}. Our model prediction for $\sigma (pp\to W_R)\times \mathcal B(W_R\to \tau_R \nu_{\tau_R})$ is given in Fig.~\ref{fig:sigmatauneutrino}, from where it follows that
the model prediction is below the ATLAS 95\% CL experimental upper bound $\sigma (pp\to W_R^1)\times \mathcal B(W_R^1\to \tau_R \nu_{\tau_R})_{exp}\lesssim 0.0035$ pb~\cite{Aaboud:2018vgh} by a factor of order of a few.

In the previous analyses we did not take into account the width of resonances. While the width (with respect to its mass $m_1$) of the KK photon $A^1$ is around $\sim$0.24, those of the other resonances  depend on the angle $\sin\theta_R$. For instance, in the range $0.35\lesssim\sin\theta_R\lesssim 0.5$ the  $Z_L^1$ width varies between 0.05 and 0.08, while those of $Z_R^1$ and $W_R^1$ are generically $\mathcal O(1)$. For the case of broad resonances, as is the case of the $Z_R^1$ and $W_R^1$ resonances, we expect that the effect of the width can affect the production cross-section (due to possible KK mode superpositions) as well as the experimental bounds (due to the absence of a clear resonance).  Recent ATLAS studies~\cite{Aaboud:2018mjh} show that bounds on the cross-sections for the case of broad resonances are affected by factors of order a few, while the cross-section predictions are also affected by similar factors.  Hence, although a detailed experimental and theoretical analysis would be necessary to determine the precise bounds on the gauge boson KK mode masses, they are expected to be of the same order as the ones shown in Figs.~\ref{fig:sigmaff} and~\ref{fig:sigmatauneutrino}. These conclusions are consistent
with the results presented in Ref.~\cite{Faroughy:2016osc} for the case of a 3~TeV vector resonance of sizable width. 

Finally there are also strong constraints on the mass of KK gluons $\mathcal G^1$ from the cross-section $\sigma(pp\to\mathcal G^1) \times \mathcal B(\mathcal G^1\to\bar tt)$ from the ATLAS experimental analysis in Ref.~\cite{Aaboud:2018mjh}. As the resonance $\mathcal G^1$ is a broad one, both the experimental results and the theoretical calculation of the production cross sections should be re-analyzed to get reliable bounds on the mass of the KK gluons. However, a simple way of relaxing the bounds is introducing brane kinetic terms for the $SU(3)$ gauge bosons, in particular in the IR brane. This theory has been analyzed in Refs.~\cite{Davoudiasl:2002ua,Carena:2002dz}, where it is shown that, even for small coefficients in front of the brane kinetic terms, the coupling of the KK modes $\mathcal G^n$ to IR localized fermions decreases very fast while the mass of the modes $m_n$ increases. Both facts going in the same directions, the bounds on KK gluons can be  easily avoided. As the strong sector does not interfere with the electroweak one $SU(2)_L\otimes SU(2)_R\otimes U(1)_X$, the presence of brane kinetic terms will not affect our mechanism for reproducing the $R_{D^{(\ast)}}$ anomaly. Moreover in the presence of brane kinetic terms for $SU(3)$ gauge bosons the flavor bounds in Sec.~\ref{sec:flavor} should be subsequently softened, an analysis that, to be conservative, we are not considering in this paper.

\section{Predictions}
\label{sec:Predictions}
In this section we will present some predictions of our theory consistent with the experimental value of $R_{D^{(\ast)}}$ and all the previously analyzed experimental constraints.

\subsection{The forward-backward asymmetry $A_{FB}^b$}
\label{sec:Zbb}
We shall study the shifts in the couplings $g_{Z_L b_{L,R} b_{L,R}}$, parametrized as 
\be
g_{Z_L b_{L,R} b_{L,R}}=g_{Z_L b_{L,R} b_{L,R}}^{SM}+\delta g_{Z_L b_{L,R} b_{L,R}}.
\ee
The shift of these couplings induce an anomalous modification of the forward-backward bottom asymmetry, conventionally defined as
\be
A_{FB}^b=\frac{3}{4}A_{LR}^e\,\left(\frac{g_{Z_Lb_Lb_L}^2-g_{Z_Lb_R b_R}^2}{g_{Z_Lb_L b_L}^2+g_{Z_l b_R b_R}^2}  \right) 
\ee
where 
\be
A_{LR}^e=\left(\frac{g_{Z_Le_Le_L}^2-g_{Z_Le_R e_R}^2}{g_{Z_Le_L e_L}^2+g_{Z_l e_R e_R}^2}  \right) \,.
\ee
The currently measured value of $\delta A_{FB}^b=\left.A_{FB}^b\right|_{\rm exp}-\left.A_{FB}^b\right|_{\rm SM}$ is given by,
\be
\delta A_{FB}^b=-0.0038\pm 0.0016,
\label{eq:AFB}
\ee
and hence $A_{FB}^b$ exhibits a $\sim$2.3~$\sigma$ anomalous departure with respect to the SM prediction~\cite{Patrignani:2016xqp}.

%
%

In our model the values of $\delta g_{Z_L b_{L} b_{L}}$ and $\delta g_{Z_L b_{R} b_{R}}$ are induced by the $Z_LZ_{L,R}^n$ mixing, in turn induced by the electroweak breaking, followed by the corresponding coupling $g_{Z_L b_{L} b_{L}}$ or $g_{Z_R b_{R} b_{R}}$~\footnote{A related analysis of the bottom forward-backward asymmetry in
models with custodial symmetry in warped extra dimensions has been performed in Ref.~\cite{Djouadi:2011aj}}, and by one-loop radiative corrections induced by the operators in Eq.~(\ref{eq:OfRtR}).
An analysis similar to that done in Sec.~\ref{sec:Ztautau} yields the expressions
\begin{align}
\delta g_{Z_L b_R b_R}&= \sum_n \left( \frac{g_R v G_3^n}{2 m_n} \right)^2\Bigg\{ \frac{r_h(\alpha)}{\cos^2\theta_R}\left[\sin^2\theta_\Sigma\left(1-\frac{2}{3}\sin^2\theta_R \right)-\frac{1}{2}\cos^2\theta_R 
\right]
\label{eq:deltagZLbRbR}
\\
+&\frac{3h_t^2}{4\pi^2}\log\frac{m_1}{m_t}
\left[\frac{2}{9}\sin^2\theta_R+\frac{1}{\cos^2\theta_R}\left(\frac{1}{2}-\frac{1}{3}\sin^2\theta_R\right)
\left(\frac{1}{2}-\frac{2}{3}\sin^2\theta_R\right)
\right]\Bigg\}
\nonumber
\end{align}
and
\begin{align}
\delta g_{Z_L b_L b_L}&= \sum_n \left( \frac{g_R v G_3^n}{2 m_n} \right)^2r_f(c_{b_L})\sin^2\theta_R \Bigg\{r_h(\alpha)\left[\frac{1}{2\sin^2\theta_L}-\frac{1}{2}+\frac{\sin^2\theta_\Sigma}{3\cos^2\theta_R} 
\right]
\label{eq:deltagZLbLbL}
\\
+&\frac{3h_t^2}{4\pi^2}\log\frac{m_1}{m_t}\left[-\frac{1}{9}+\frac{1}{6\cos^2\theta_R}\left(\frac{1}{2}-\frac{2}{3}\sin^2\theta_R  \right)  \right]\Bigg\}\nonumber
\end{align}
 \begin{figure}[htb]
\centering
\includegraphics[width=7.5cm]{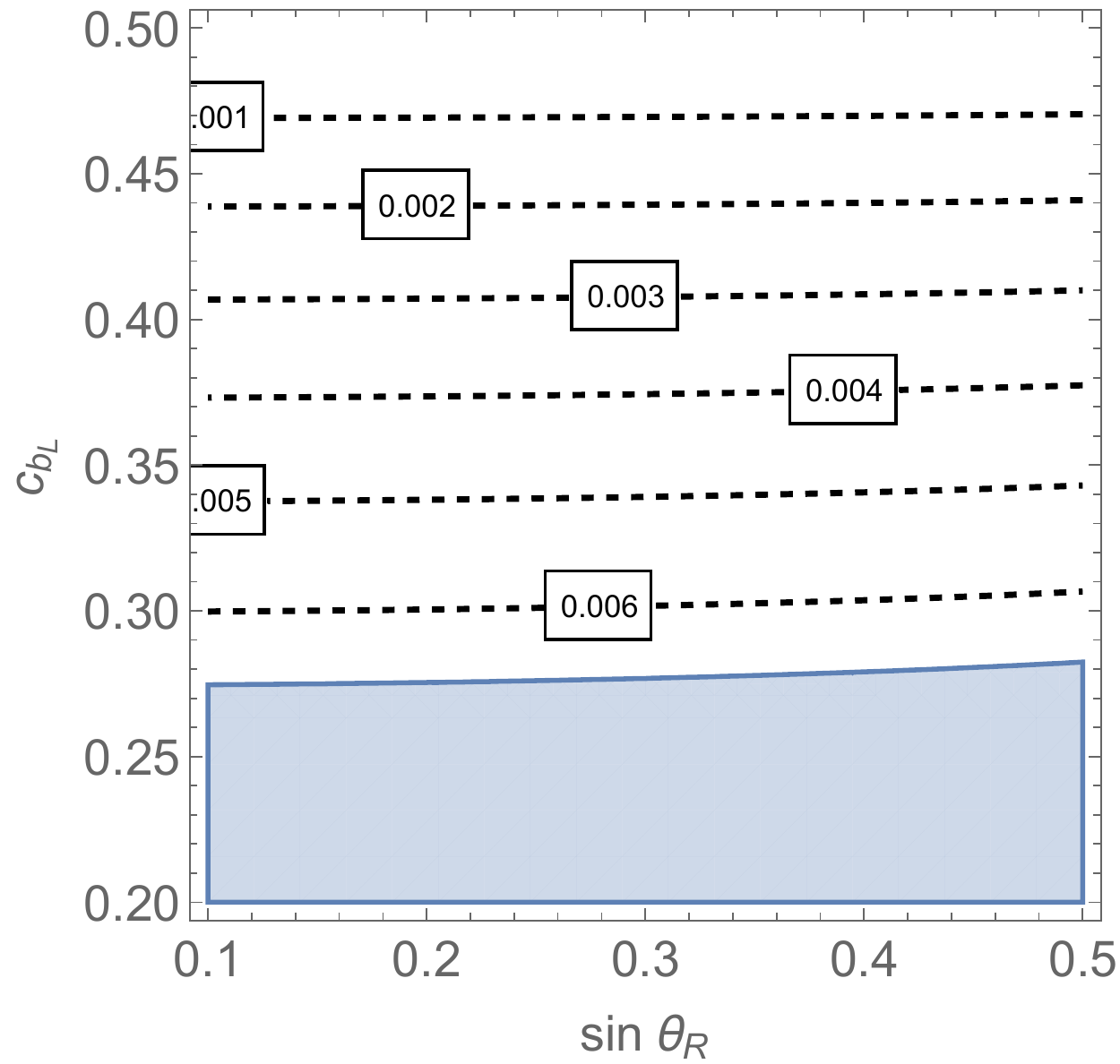} 
\includegraphics[width=7.5cm]{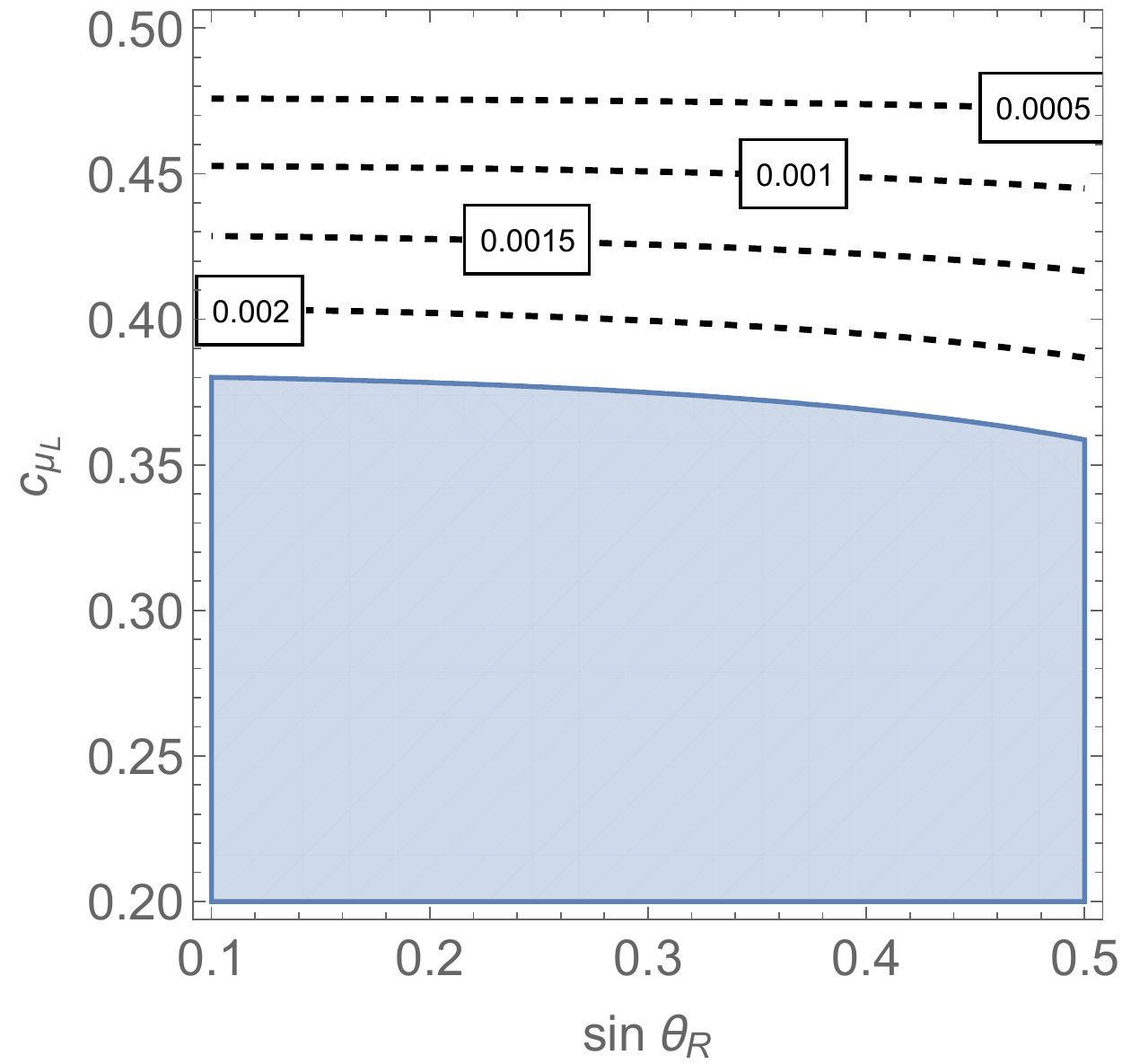} 
\caption{\it Left panel: Contour lines of $\delta g_{Z_Lb_Lb_L}$ in the plane $(\sin\theta_R,c_{b_L})$ where we have fixed $\sin\theta_\Sigma=0.72$. The white region is allowed by electroweak precision data at the 95\% CL. Right panel: The same for $\delta g_{Z_L\mu_L\mu_L}$.
}
\label{fig:deltagZLbL}
\end{figure} 
where, again, the first lines in Eqs.~(\ref{eq:deltagZLbRbR}) and (\ref{eq:deltagZLbLbL}) are the contributions from the gauge bosons KK modes through 
mixing effects, and the second lines come from the contribution of the radiative corrections induced by the operators 
$$\mathcal O_{b_{R, L}t_R}=(\bar b_{R,L }\gamma^\mu b_{R, L})(\bar t_R\gamma_\mu t_R)\ .$$

Finally, the modification of the left-handed and right-handed bottom couplings to the $Z$ gauge boson induce 
a modification of $A_{FB}^b$ which, at linear order in $\delta g_{Z b_{L,R} b_{L,R}}$ is given by 
\be
\delta A_{FB}^b=-0.183 \,\delta g_{Z_L b_R b_R}-0.033\, \delta g_{Z_L b_L b_L} .
\ee
 \begin{figure}[htb]
\centering
\includegraphics[width=10cm]{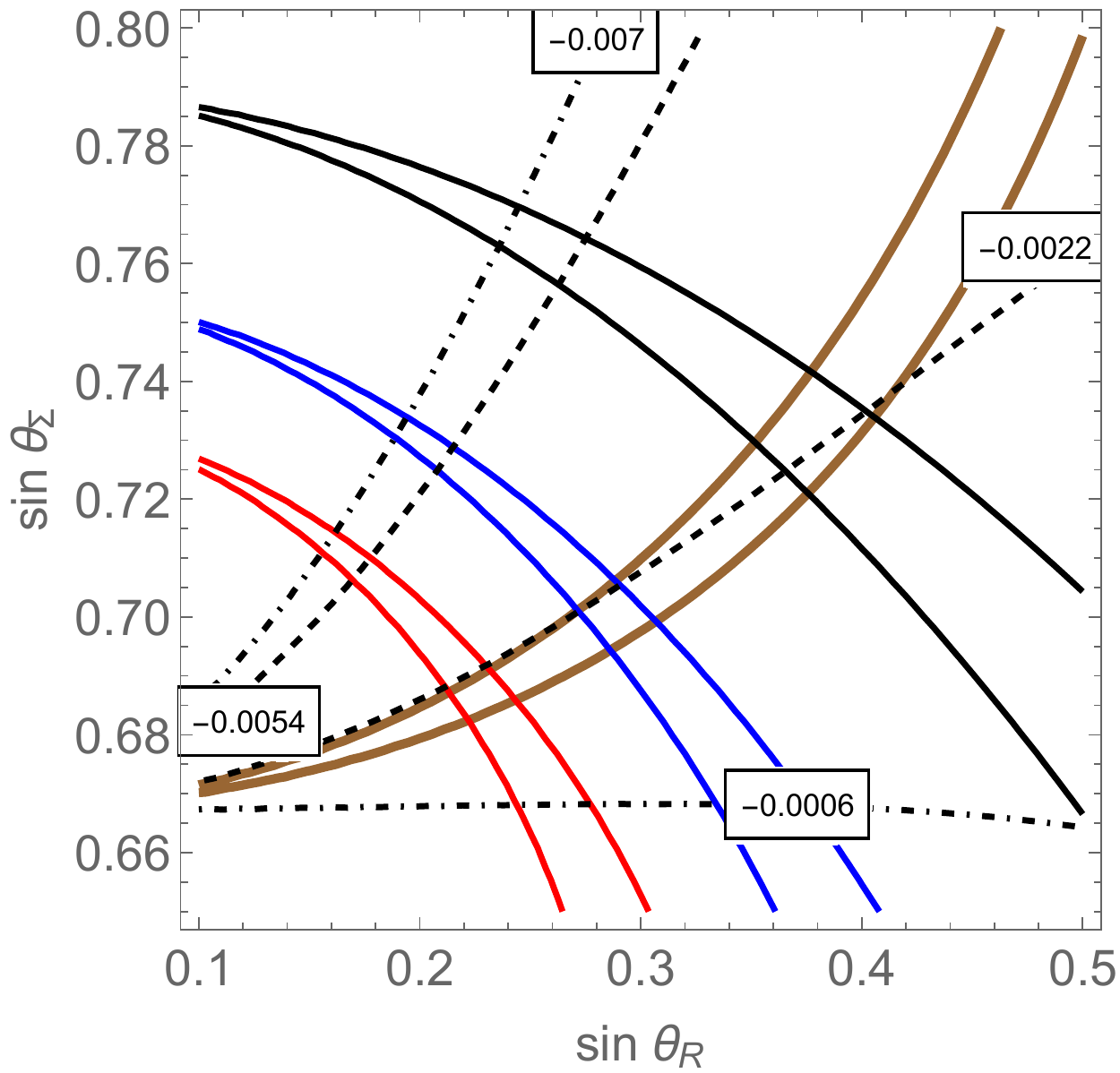} 
\caption{\it The region between the solid lines is allowed by  $\delta g_{Z_L\tau_R\tau_R}$ (brown lines) and by $T$ for  $m_{1}=3$ TeV and $\tan\beta=1$ (black lines) $\tan\beta=3$ (blue lines) and $\tan\beta=5$ (red lines). Region between dashed (dot-dashed) lines encompasses the 1~$\sigma$ (2~$\sigma$) interval for the anomaly in $A_{FB}^b$.
}
\label{fig:gZbR}
\end{figure} 
The shift $\delta g_{Z_Lb_Lb_L}$ is constrained by electroweak precision data, to be~\cite{Falkowski:2017pss}
\be
\delta g_{Z_Lb_Lb_L}=(3.3\pm 1.7)\times 10^{-3}\,.
\label{eq:ZLbLbound}
\ee
The region (\ref{eq:ZLbLbound}) constrains the available values of $c_{b_L}$, as shown in the left panel of Fig.~\ref{fig:deltagZLbL}, where we have fixed $\sin\theta_\Sigma=0.72$ and where the shaded area is excluded at the 95\% CL.

After fixing the condition to fit $R_{D^{(\ast)}}$, and using e.g.~the value $c_{b_L}=0.35$, for which $\delta g_{Z_L b_L b_L}\simeq 4.7\times 10^{-3}$, we find that the 1~$\sigma$ (2~$\sigma$) experimental value (\ref{eq:AFB}) is obtained between the dashed (dot-dashed) lines in Fig.~\ref{fig:gZbR}, implying that the anomalous value of $A_{FB}^b$ remains consistent with the explanation of the $R_{D^{(\ast)}}$ anomaly, and the rest of electroweak and LHC constraints, for the parameter region near $\tan\beta=2\pm 1$, $\sin\theta_R\simeq 0.32 \pm 0.08$ and $\sin\theta_\Sigma\simeq 0.72 \pm 0.02$. Observe,
however, that $\tan\beta$ close to one demands large values of the top-quark Yukawa coupling.  As it is clear from Fig.~\ref{fig:gZbR}, for somewhat larger values of $\tan\beta$ the corrections to the right-handed bottom coupling allow to reduce the current 2.3~$\sigma$ anomaly on $A_{FB}^b$ into a value that is about 1~$\sigma$ away from the central experimental value.

Observe that this custodial symmetry model differs from the results obtained in an abelian gauge symmetry extension of the SM, where an explanation of the forward-backward asymmetry 
demands the extra gauge bosons to be light, with masses below about 150~GeV, in order to induce small corrections to the $T$ parameter~\cite{Liu:2017xmc}.

\subsection{The processes ${B}\to {K}\nu\nu$ and $B^+ \to K^+ \tau^+ \tau^-$}
The $R_{D^{(\ast)}}$ anomaly can in principle induce a large production in the process ${B}\to {K}\bar\nu\nu$, i.e.~$b\to s\bar\nu\nu$, mainly induced by the RH neutral current Lagrangian~\footnote{Notice that $g_{Z_L\nu_R\nu_R}=g_{A\nu_R\nu_R}=0$ and hence no $Z_L^n$ or $A^n$ mediated processes occur.}
\be
\mathcal L=\frac{g_R}{\cos\theta_R}\sum_{n=1}^\infty \left\{ (V_{d_R}^\dagger)_{23}\,g_{Z_Rd_Rd_R}\,G_3^n\,(\bar s_R\, \slashed{Z}_R^n b_R)+g_{Z_R\nu_R\nu_R}\,G_3^n\,(\bar\nu_R \slashed{Z}_R^n\nu_R)
\right\}
\ee
where the couplings of $Z_R$ to RH quarks and leptons are given in Eq.~(\ref{eq:gZR}).
After integrating out the KK modes we get the effective Lagrangian
\begin{align}
\mathcal L_{eff}^{\nu\nu}&=-g_{Z_Rd_Rd_R}\, g_{Z_R\nu_R\nu_R}\, \frac{1}{2\cos^2\theta_R}(V_{d_R}^\dagger)_{23}
\sum_n\left(\frac{g_R\, G_3^n}{m_n}\right)^2
\ (\bar s_R\gamma^\mu b_R)(\bar\nu\gamma_\mu(1+\gamma_5)\nu)\nonumber\\
&\equiv -\frac{4G_F}{\sqrt{2}}\, V_{tb}V^*_{ts}\frac{\alpha_{EM}}{4\pi}\, C_{\nu\nu}\ (\bar s_R\gamma^\mu b_R)(\bar\nu\gamma_\mu(1+\gamma_5)\nu)
\end{align}
 \begin{figure}[htb]
\centering
\includegraphics[width=7.5cm]{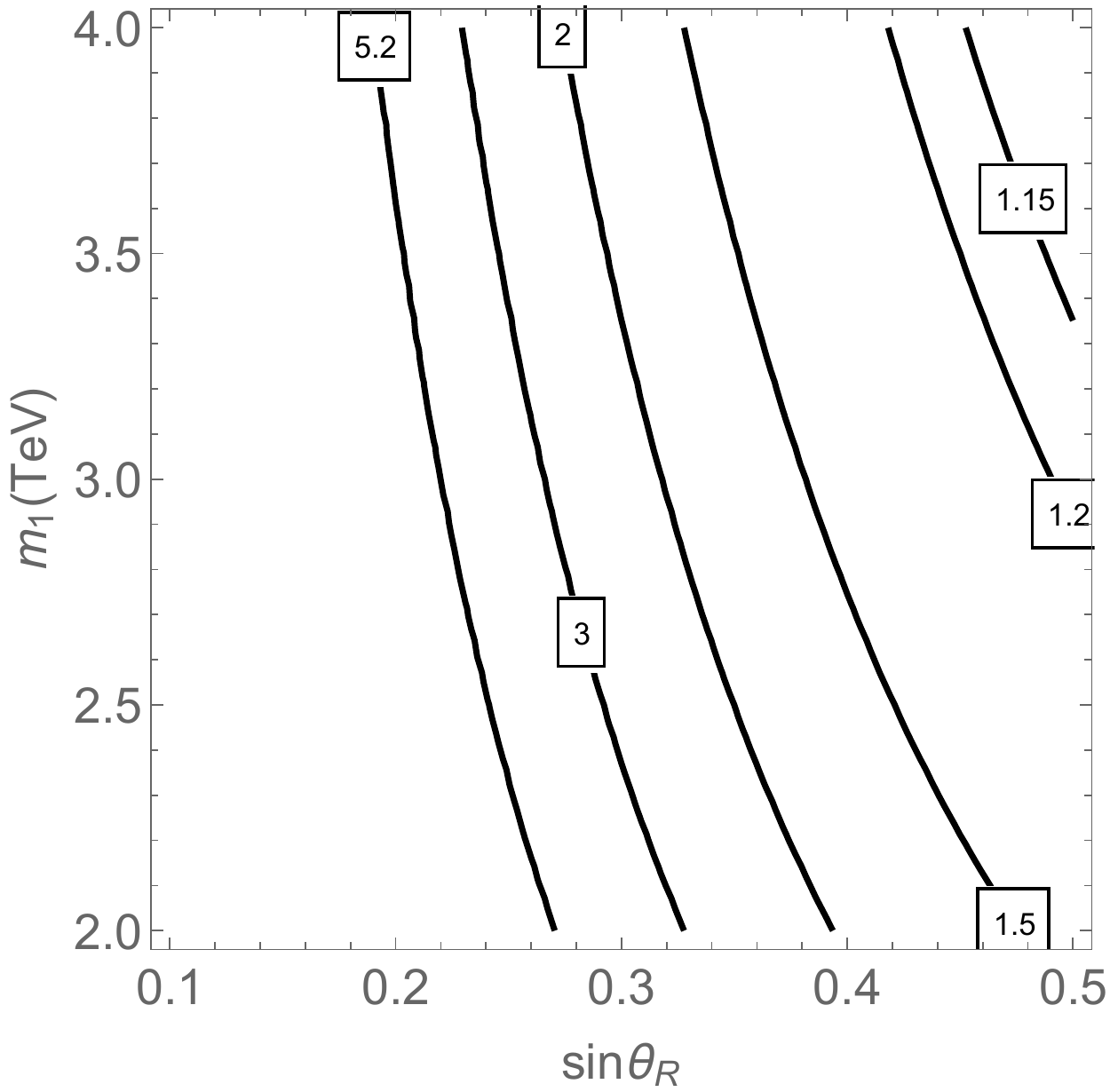} 
\includegraphics[width=8cm]{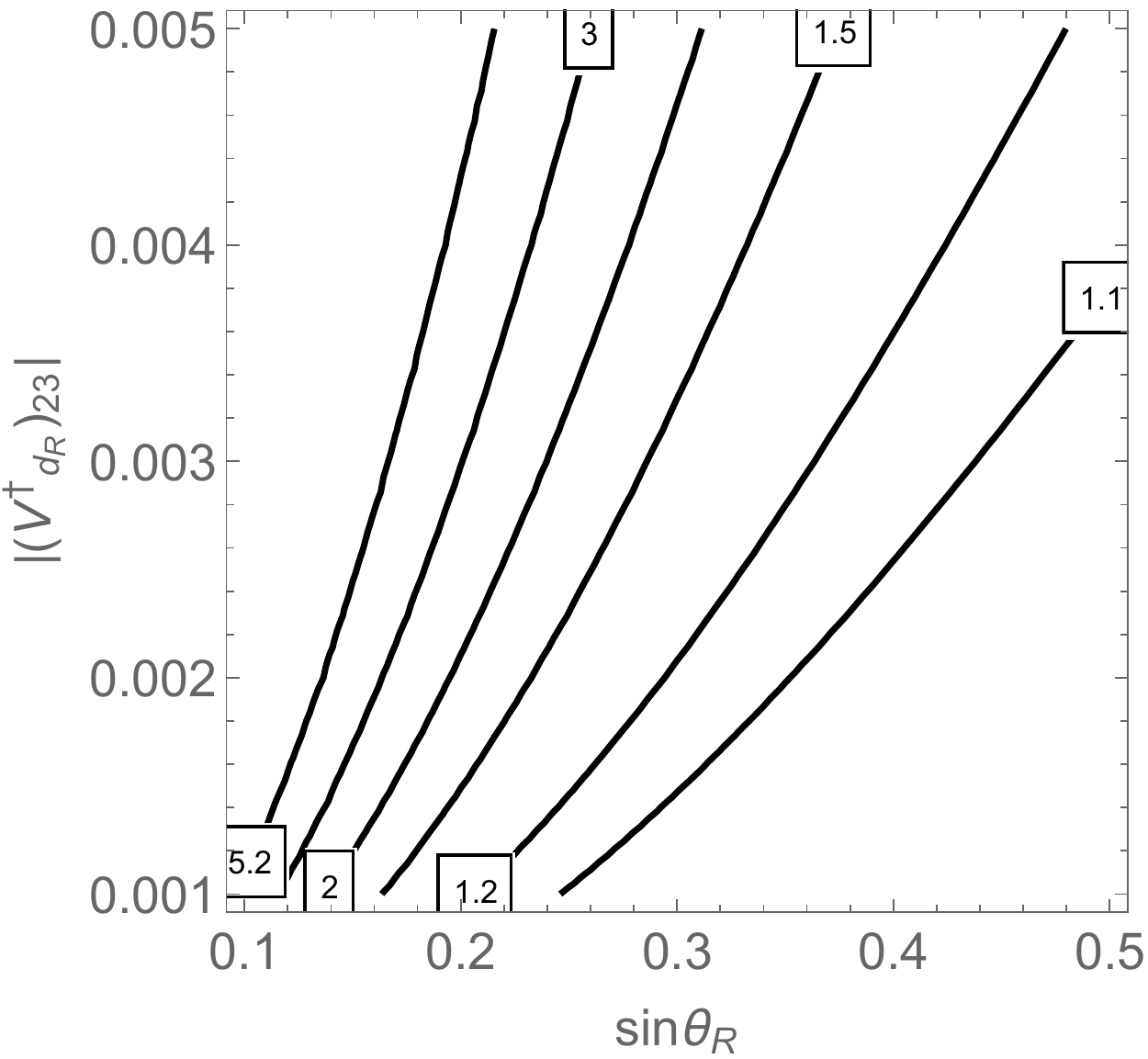} 
\caption{\it  Left panel: Contour lines of constant $R_K^{\nu\nu}$ where we have used the upper bound on  $(V^\dagger_{d_R})_{23}= 0.005$ obtained by the condition given in Eq.~(\ref{eq:Cbdbs}) . Right panel: Contour lines of $R_K^{\nu\nu}$ for $m_{1}=3$ TeV.
}
\label{fig:VdR}
\end{figure} 
where we are normalizing $\mathcal B(B\to K\nu_R\nu_R)$ to the SM value of 
$\sum_\ell\mathcal B(B\to K\nu_\ell\nu_\ell)$, and the Wilson coefficient $C_{\nu\nu}$ is given by
\be
C_{\nu\nu}=\frac{1}{2\cos^2\theta_R}\left(\frac{1}{2}-\frac{1}{3}\sin^2\theta_R\right)\frac{4\pi}{\alpha_{EM}}\, (V^\dagger_{d_R})_{23}  \frac{1}{V_{cb}}\sum_n\left( \frac{g_R v G_3^n}{2m_n} \right)^2  
\ee
and where we have used that in the Wolfenstein parametrization $V_{cb} = - V_{ts} = A\lambda^2$, and $V_{tb} = 1$.

Now we can write the ratio
\be
R_K^{\nu\nu}=\frac{\mathcal B(B\to K\nu\nu)}{\mathcal B(B\to K\nu\nu)_{SM}}=1+\frac{1}{3}\,
\frac{|C_{\nu\nu}|^2}{|C_{\nu\nu}^{SM}|^2}\simeq 1+0.008\ |C_{\nu\nu}|^2 ,
\ee
where we have used the SM prediction $C_{\nu\nu}^{SM}\simeq -6.4$~\cite{Buras:2014fpa}. Using the experimental bound 
$R_K^{\nu\nu}<5.2$ at the 95\% CL~\cite{Grygier:2017tzo}, one finds the bound $|C_{\nu\nu}|\lesssim 23$. However, after imposing the constraints coming from the flavor condition (\ref{eq:Cbdbs}) on the matrix element $(V^\dagger_{d_R})_{23}$, one easily obtains values that are well below the experimental bound, particularly for values of
$\sin\theta_R > 0.2$. This is shown in the left panel of Fig.~\ref{fig:VdR}, where we plot contours of constant $R_K^{\nu\nu}$ in the plane $(\sin\theta_R,m_{1})$ after using the bound for $(V^\dagger_{d_R})_{23}$ in Eq.~(\ref{eq:Cbdbs}).  Lower values of $R_K^{\nu\nu}$ may be obtained for smaller values of $\sin\theta_R$
by using the freedom on the value of $(V^\dagger_{d_R})_{23}$, as shown in the right panel of Fig.~\ref{fig:VdR}, where we plot $R_K^{\nu\nu}$ in the plane $\left(\sin\theta_R,(V^\dagger_{d_R})_{23}\right)$ after fixing $m_{1}=3$ TeV. 

\begin{figure}[htb]
\centering
\includegraphics[width=9cm]{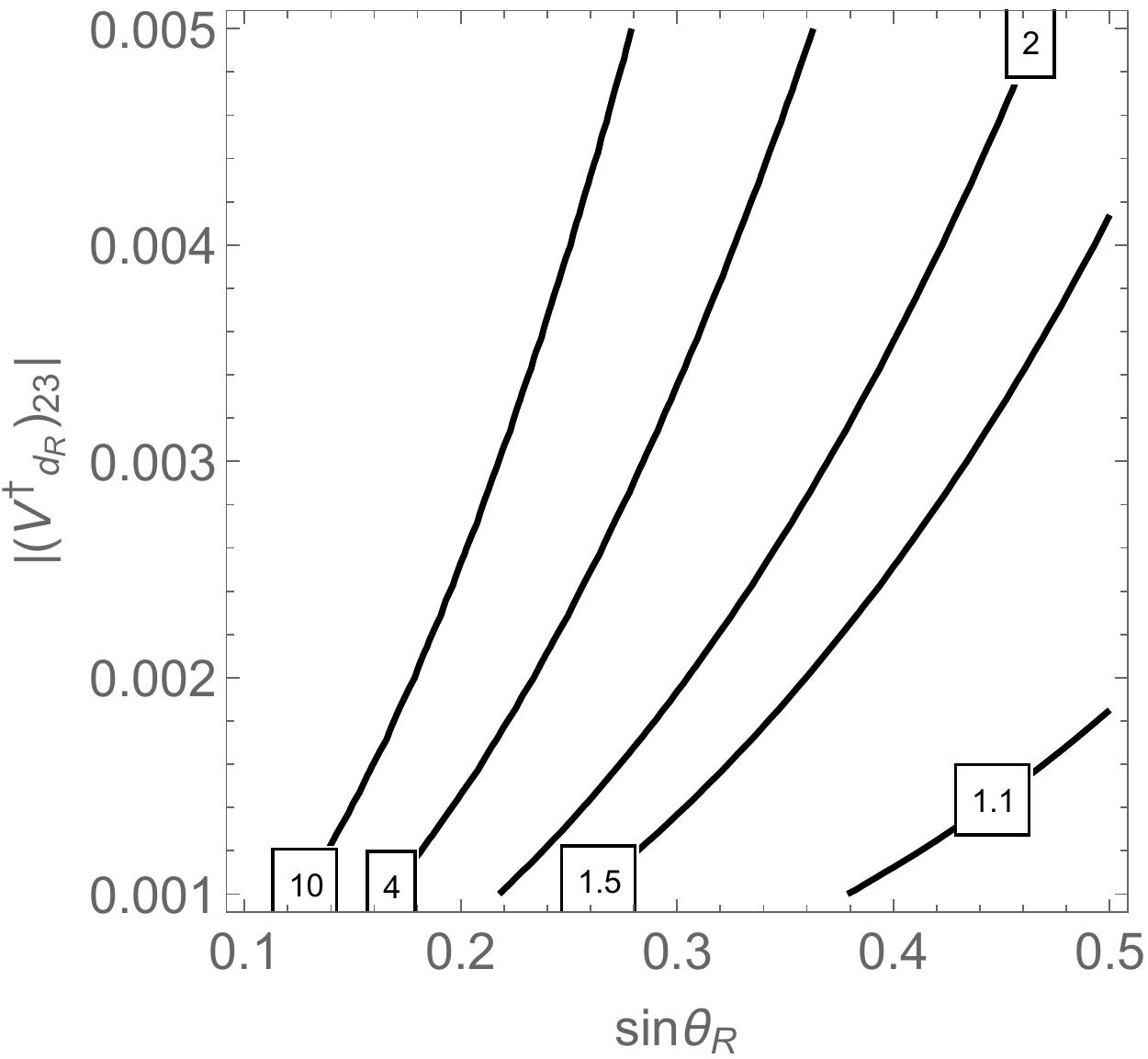} 
\caption{\it  Contour lines of $R_{K}^\tau$ in the plane $[\sin\theta_R,|(V^\dagger_{d_R})_{23}|]$, for values of $(V^\dagger_{d_R})_{23}$ consistent with the flavor constraints.}
\label{fig:tautau}
\end{figure}

This model predicts a strong $\tau\tau$ production in the observable
\be
R_{K}^\tau=\frac{\mathcal B(B^+\to K^+\tau\tau)}{\mathcal B(B^+\to K^+\tau\tau)_{SM}} \,.
\ee  
In our model this observable is dominated by the Wilson coefficient $C_{RR}^\tau$ such that 
\be
R_{K}^\tau\simeq 1+\left|\frac{C_{RR}^\tau}{C_{LL}^{SM}}\right|^2
\ee  
where
\begin{align}
C_{RR}^\tau &= -\frac{8\pi}{\alpha_{EM}}\sum_n\left(\frac{g_R v G_3^n}{2m_n}  \right)^2 \frac{(V_{d_R}^\dagger)_{23}}{V_{ts}^\ast}  \nonumber \\
&\quad \times \left[ \frac{1}{3}\sin^2\theta_R+\frac{1}{\cos^2\theta_R}\left(\frac{1}{2}-\frac{1}{3}\sin^2\theta_R\right)\left(\frac{1}{2}-\sin^2\theta_R\right) 
\right]  \,.
\end{align}
Contour lines of constant $R_{K}^\tau$ are presented in Fig.~\ref{fig:tautau} for $m_1 =3$~TeV.
The results are widely consistent with present experimental bounds from the BaBar Collaboration~\cite{TheBaBar:2016xwe} which yield the 90\% CL upper bound, $R_{K}^\tau<10^4$.

\subsection{$R_{K^{(\ast)}}$ }

One of the general applications of our theory is that it generically predicts a value of $R_{K^{(\ast)}}$ 
\begin{equation}
R_{K^{(*)}} = \frac{{\rm BR}(B \to K^{(*)} \mu^+ \mu^-)}{{\rm BR}(B \to K^{(*)} e^+ e^-)},
\end{equation}
which can easily differ from its SM prediction~\cite{Aaij:2014ora,Aaij:2017vbb}. The general effective operator Lagrangian is written as
\be
\mathcal L_{eff}=\frac{4G_F}{\sqrt{2}}\frac{\alpha_{EM}}{4\pi}V^\ast_{ts}V_{tb}\sum_i C_i \mathcal O_i .
\ee
We will find it convenient to work  in the chiral basis for the operators $\mathcal O_i$ such that operators
\be
\mathcal O_{\chi\chi^\prime}=(\bar s_\chi\gamma^\mu b_\chi)(\bar\ell_{\chi^\prime}\gamma_\mu\ell_{\chi^\prime})
\ee
with chiralities $\chi,\chi^\prime=\{L,R\}$, have Wilson coefficients defined as $C_{\chi\chi^\prime}^\ell\equiv C_{\chi\chi^\prime}^{SM}+\Delta C_{\chi\chi^\prime}^\ell$~\footnote{The relation with the usual non-chiral basis, $C_{9,10}^{(')}$,~\cite{Buchalla:1995vs} is given by: $C_{LL}=C_9-C_{10}$, $C_{LR}=C_9+C_{10}$, $C_{RL}=C^\prime_9-C^\prime_{10}$ and $C_{RR}= C^\prime_9+C^\prime_{10}$.}.
The SM predictions are given by
\be
C^{SM}_{LL}\simeq 8.4,\ C^{SM}_{RL}\simeq C^{SM}_{LR}\simeq C^{SM}_{RR}\simeq 0
\ee
while $\Delta C^{\ell}_{\chi\chi^\prime}$ are the contributions to the Wilson coefficients coming from New Physics.

The prediction of $R_{K^{(\ast)}}$ is given by
\be
R_{K^{(\ast)}}=\frac{|C^\mu_{LL}+C^\mu_{LR} \pm C^\mu_{RL} \pm C^\mu_{RR}|^2+|C^\mu_{LR}-C^\mu_{LL} \pm C^\mu_{RR} \mp C^\mu_{RL}|^2}{
|C^e_{LL}+C^e_{LR} \pm C^e_{RL} \pm C^e_{RR}|^2+|C^e_{LR}-C^e_{LL} \pm C^e_{RR} \mp C^e_{RL}|^2}
\label{eq:RK}
\ee
where the upper signs correspond to $R_K$ and the lower signs to $R_{K^*}$ and
we have assumed that the polarization of the $K^*$ is close to $p =1$, what is a good approximation in the relevant $q^2$ region
associated with the $R_{K^*}$ measurement~\cite{Hiller:2014ula}. The above
equation, Eq.~(\ref{eq:RK}),  shows the well known correlation (anti-correlation) of the corrections to $R_K$ and $R_{K^*}$ associated
to the left- (right-) handed currents. Therefore, considering the fact that both $R_K$ and $R_{K^*}$ are suppressed with respect to the SM
values, this leads to a preference of new physics effects involving left-handed currents.

The experimental value of $R_{K^{(\ast)}}$ departs from the SM prediction $R_{K^{(\ast)}}\simeq 1$~\cite{Bobeth:2007dw} by around 2.5~$\sigma$. 
Moreover global fits~\cite{Crivellin:2015era,Descotes-Genon:2015uva,Descotes-Genon:2016hem,Hurth:2016fbr,Altmannshofer:2017fio,Capdevila:2017bsm,Mahmoudi:2018qsk} to a number of observables, including the branching ratios for $B\to K^\ast\ell\ell$, $B_s\to\phi\mu\mu$, and $B_s\to \mu\mu$, favor a solution where $\Delta C^\mu_{LL}<0$ while $\Delta C^\mu_{RL}\simeq \Delta C^\mu_{LR}\simeq \Delta C^\mu_{RR}\simeq 0$, and $\Delta C^e_{\chi,\chi^\prime}\simeq 0$ for $\chi,\chi^\prime=\{L,R\}$. 

In fact, in our model, for
\be
c_{e_{L,R}}\gtrsim 1/2,\quad c_{\mu_R}\gtrsim 1/2
\ee
it turns out that $C^e_{\chi\chi^\prime}\simeq C^{SM}_{\chi\chi^\prime}$ and $\Delta C^\mu_{LR}\simeq \Delta C^\mu_{RR}\simeq 0$~\footnote{Or, in the usual basis language,~$\Delta C_9^\mu=-\Delta C_{10}^\mu$ and $\Delta C^{\prime \mu}_{9,10}\simeq 0$.}. On the other hand the prediction for $\Delta C^\mu_{LL}$  is given by
\begin{align}
\Delta C^\mu_{LL}&=-\frac{8\pi}{\alpha_{EM}}\sum_n\left( \frac{g_Rv G_3^n}{2m_n}\right)^2 r_f(c_{b_L})r_f(c_{\mu_L})\cdot\nonumber\\
&\sin^2\theta_R\left[\left(\frac{1}{2\sin^2\theta_L}-\frac{1}{3} \right)\left(\frac{1}{2}-\sin^2\theta_L  \right)+\frac{1}{3}\cos^2\theta_L-\frac{1}{12}\frac{\sin^2\theta_R}{\cos^2\theta_R}
\right]  \,,
\end{align}
where the first, second and third terms inside the square bracket comes from the contribution of the $Z_L^n$, $A^n$ and $Z_R^n$ KK modes, respectively, and
we are assuming~\cite{Cabrer:2011qb} that $V_{u_L}\simeq 1$ and $V_{d_L}\simeq V$, the CKM matrix. Similarly, the prediction for $\Delta C^\mu_{RL}$
is given by
\begin{align}
\Delta C^\mu_{RL}&=-\frac{8\pi}{\alpha_{EM}}\sum_n\left( \frac{g_Rv G_3^n}{2m_n}\right)^2 r_f(c_{\mu_L})\,\left[(V^\dagger_{d_R})_{23}/V^\ast_{ts} \right]\cdot \\
& 
\sin^2\theta_R\left[\frac{1}{3}\left(-\frac{1}{2}+\sin^2\theta_L \right)+\frac{1}{3}\cos^2\theta_L+\frac{1}{2\cos^2\theta_R}\left(-\frac{1}{2}+\frac{1}{3}\sin^2\theta_R  \right)
\right] .
\nonumber
\end{align}
Observe that the combined contribution to $\Delta C^\mu_{LL}$ from the $Z_L^n$ and $A^n$ KK modes is considerably larger than the one from the $Z_R^n$ KK modes.
\begin{figure}[htb]
\centering
\includegraphics[width=7.9cm]{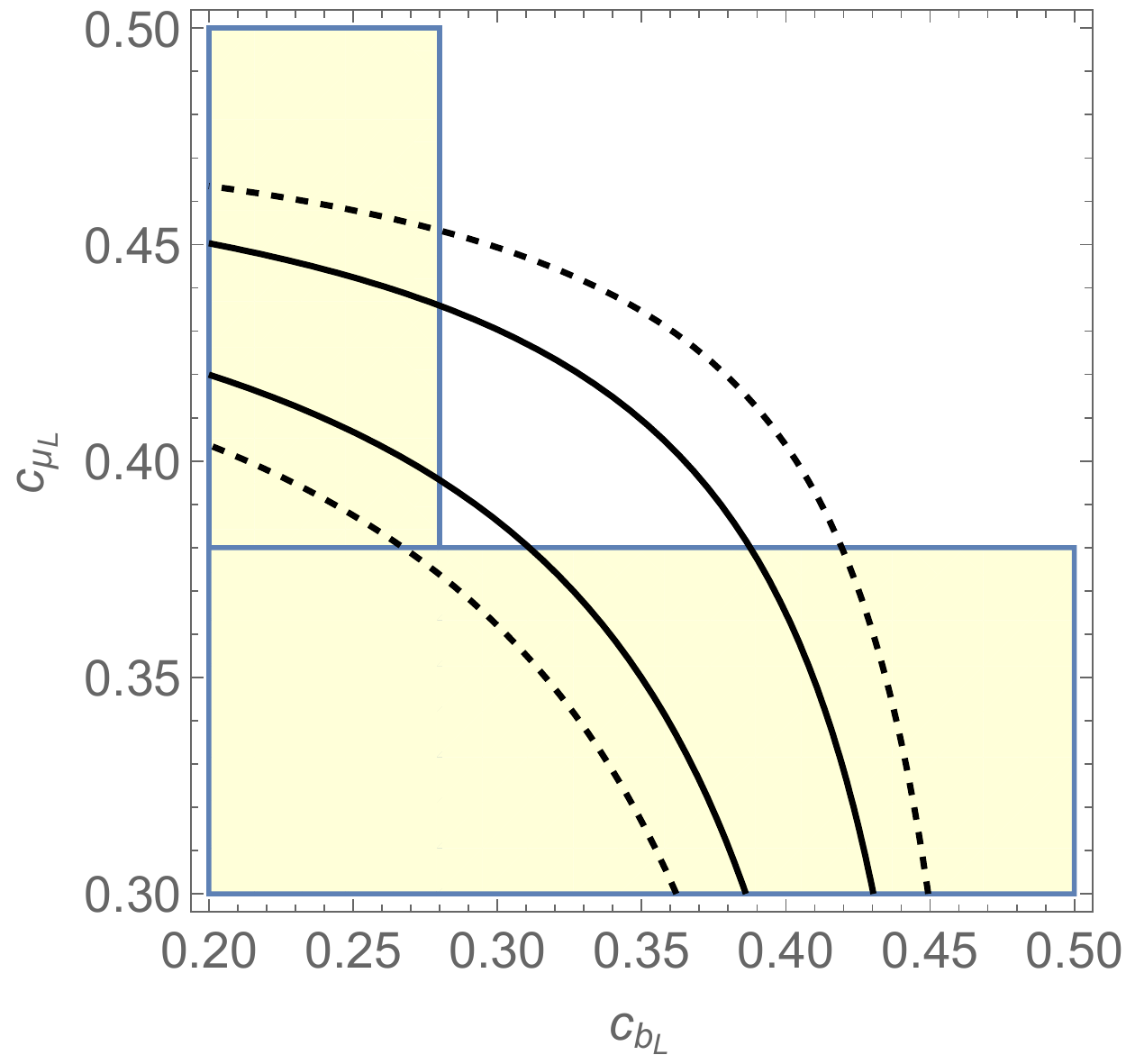} 
\includegraphics[width=7.9cm]{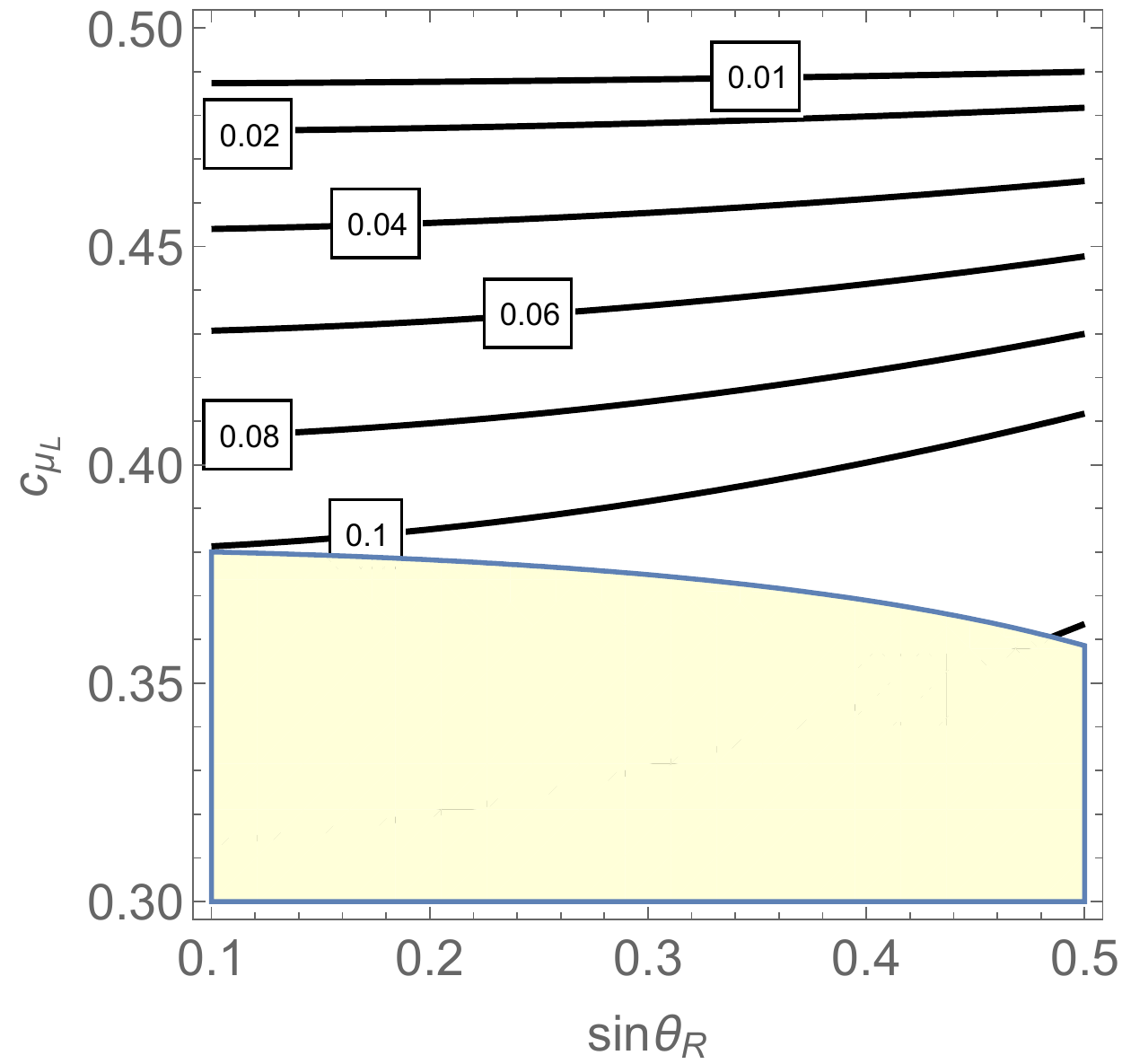} 
\caption{\it  Left panel: Contour lines of constant $\Delta C_{LL}^\mu$ in the plane $(c_{b_L},c_{\mu_L})$ using $V_{d_L}\simeq V$,  at 1~$\sigma$ (solid lines) and 2~$\sigma$ (dashed lines) level as defined
in Eq.~(\ref{eq:CLL}). The value of $\sin\theta_R = 0.35$ has been fixed. Right panel:
Contour lines of constant $\Delta C_{RL}^\mu$ in the plane $(\sin\theta_R,c_{\mu_L})$
after selecting the upper bound  on $|(V^\dagger_{d_R})_{23}|\simeq 0.005$ obtained in Eq.~(\ref{eq:Cbdbs}). In both panels the shaded yellow area corresponds to the excluded area
obtained in Fig.~\ref{fig:deltagZLbL}.
}
\label{fig:RK}
\end{figure} 
Recent global fits to experimental data~\cite{Altmannshofer:2017fio} yield the 1~$\sigma$ (2~$\sigma$) prediction
\be
\Delta C_{LL}^\mu \subset [-1.66,-1.04]_{1\sigma},\  [-1.98,-0.76]_{2\sigma}
\label{eq:CLL}
\ee
which constitutes a $\sim$~4.8~$\sigma$ deviation with respect to the SM prediction. On the other hand $C_{RL}^\mu$ has to be small and in fact the global fit yields~\cite{Altmannshofer:2017fio}
\be
\Delta C_{RL}^\mu \subset [-0.04,0.36]_{1\sigma},\  [-0.24,0.56]_{2\sigma}
\label{eq:CRL}
\ee
which only depart $\sim$0.8~$\sigma$ from the SM prediction. 

The left panel of Fig.~\ref{fig:RK} shows the $1~\sigma$ (solid lines) and $2~\sigma$ (dashed lines) contours of  $\Delta C^\mu_{LL}$ in the plane $(c_{b_L},c_{\mu_L})$, where we have fixed $\sin\theta_R=0.35$. 
The values of $c_{b_L}$ and $c_{\mu_L}$ are mainly constrained from $\delta g_{Z_Lb_Lb_L}$, as given in Eq.~(\ref{eq:deltagZLbLbL}), and plotted in the left panel of Fig.~\ref{fig:deltagZLbL},
and from $\delta g_{Z_L\mu_L\mu_L}$ as given by
\begin{align}
\delta g_{Z_L \mu_L \mu_L}&= \sum_n \left( \frac{g_R v G_3^n}{2 m_n} \right)^2 r_f(c_{\mu_L})\sin^2\theta_R\Bigg\{r_h(\alpha)\left[\frac{1}{2\sin^2\theta_L}-\frac{1}{2}-\frac{\sin^2\theta_\Sigma}{\cos^2\theta_R} 
\right]\nonumber\\
+&\frac{1}{12\cos^2\theta_R}\frac{3h_t^2}{4\pi^2}\log\frac{m_1}{m_t} 
\Bigg\} \ ,\label{eq:deltagZLmuL}
\end{align}
where, again, the first line in Eq.~(\ref{eq:deltagZLmuL}) denote the contributions from the gauge bosons KK modes through the mixing and the second line denote those from the radiative corrections induced by the operators 
$$\mathcal O_{\mu_Lt_R}=(\bar \mu_{L }\gamma^\mu \mu_{L})(\bar t_R\gamma_\mu t_R)\ .$$

The prediction for $\delta g_{Z_L\mu_L\mu_L}$ is plotted in the right panel of Fig.~\ref{fig:deltagZLbL}, where we also have fixed $\sin\theta_\Sigma=0.72$, and where the white region is allowed at the 95\% CL given the fitted value to experimental data~\cite{Falkowski:2017pss}
\be
\delta g_{Z_L\mu_L\,u_L}=(0.1\pm 1.2)\times 10^{-3}.
\ee
Moreover, from Fig.~\ref{fig:deltagZLbL} at 95\% CL, $c_{b_L}\gtrsim 0.28$ and $c_{\mu_L}\gtrsim 0.38$, independently on the value of $\sin\theta_R$. The forbidden regions in Fig.~\ref{fig:RK} are represented by shaded light-yellow areas. 

The prediction for $C^\mu_{RL}$ is shown in the right panel of Fig.~\ref{fig:RK} in the plane $(\sin\theta_R,c_{\mu_L})$ where we are already using the upper bound on $(V^\dagger_{d_R})_{23}$ from flavor observables, while the shaded region is excluded by $\delta g_{Z_L\mu_L\mu_L}$. We see that the values of $\Delta C^\mu_{RL}$ in the region defined by Eq.~(\ref{eq:RK}) are always $\lesssim\mathcal O(0.1)$ and hence in accordance with the global fits, Eq.~(\ref{eq:CRL}).

\section{Conclusions}
\label{conclusions}

The experimental measurements of $R_{D^{(*)}}$ show  significant deviations from the SM values,
a surprising result due to the tree-level nature of this process in the SM.  Possible resolutions of 
this anomaly face significant constraints from the excellent agreement of flavor physics observables
with the values predicted within the SM.  In this work, we have presented an explicit realization of the solution to the 
$R_{D^{(*)}}$ anomaly based on  the contribution of right-handed currents of quarks and leptons to this process. The model is based on
the embedding of the SM in warped space, with a bulk gauge symmetry $SU(2)_L \otimes SU(2)_R \otimes U(1)_{B-L}$,
with third-generation right-handed quarks and leptons localized on the infrared-brane, ensuring a large coupling of these modes
to the charged gauge boson $W^{n}_R$ KK-modes. 

The right-handed $SU(2)_R$ gauge boson KK-modes provide the necessary contribution to $R_{D^{(*)}}$, due to
relevant mixing parameters in the right-handed up-quark sector. This may be done without inducing
large contributions to the $B$-meson invisible decays, or the $B$-meson mixings, since these observables strongly
depend on  the down-quark right handed mixing angles, which do not affect $R_{D^{(*)}}$ in any
significant way within this framework. The mass of the lightest KK-mode tends to be of about a few~TeV,
and it is in natural agreement with current LHC constraints.  

An important assumption within this model is that there is no mixing in the lepton sector. This can
be ensured with appropriate symmetries, that must be (softly) broken in order to allow the proper 
neutrino mixing. We have presented a scenario, based on symmetries and a double seesaw mechanism,
 that allows for a proper description of the lepton sector of the model.  The origin of the new parameters in the
 lepton sector remains, however, as one of the most challenging aspects of these (and many) scenarios. 
Aside of this question, beyond providing a resolution to the $R_{D^{(\ast)}}$ anomaly, this model also provides a solution 
of the hierarchy problem, has an explicit custodial symmetry that implies small corrections to the precision electroweak 
observables, and allows a solution to the $R_{K^{(\ast)}}$ anomalies mainly via the contribution of the $SU(2)_L$ KK modes. Moreover, the proposed model naturally predicts an anomalous value of the forward-backward asymmetry $A_{FB}^b$, as implied by LEP data, driven by the $Z\bar b_R b_R$ coupling.

\section*{\sc Acknowledgments}
This manuscript has been authored by Fermi Research Alliance, LLC
under Contract No. DE-AC02-07CH11359 with the U.S. Department of
Energy, Office of Science, Office of High Energy Physics. The United
States Government retains and the publisher, by accepting the article
for publication, acknowledges that the United States Government
retains a non-exclusive, paid-up, irrevocable, world-wide license to
publish or reproduce the published form of this manuscript, or allow
others to do so, for United States Government purposes. Work at
University of Chicago is supported in part by U.S. Department of
Energy grant number DE-FG02-13ER41958. Work at ANL is supported in
part by the U.S. Department of Energy under Contract
No. DE-AC02-06CH11357. The work of E.M. is supported by the Spanish
MINEICO and European FEDER funds grant number FIS2017-85053-C2-1-P,
and by the Junta de Andaluc\'{\i}a grant number FQM-225. The research
of E.M. is also supported by the Ram\'on y Cajal Program of the
Spanish MINEICO under Grant RYC-2016-20678. The work of M.Q. is partly
supported by Spanish MINEICO under Grant CICYT-FEDER-FPA2014-55613-P
and FPA2017-88915-P, by the Catalan Government under Grant
2017SGR1069, and by the Severo Ochoa Excellence Program of MINEICO
under Grant SEV-2016-0588. M.Q. would like to thank the Argonne
National Laboratory and the Fermi National Accelerator Laboratory,
where part of this work has been done, for hospitality and the Argonne
National Laboratory for financial support.  We would like to thank
Antonio Delgado, Admiri Greljo, Da Liu, J. Liu, E. Stamou, N.R. Shah,
David Shih and Jure Zupan for useful discussions, and specially
X. Wang, for her help in computing the LHC KK gauge boson cross
sections.  E.M., M.Q. and C.W. would like to thank the Mainz Institute
for Theoretical Physics, and M.C. and C.W.  the Aspen Center for
Physics, which is supported by National Science Foundation grant
PHY-1607611, for the kind hospitality during the completion of this
work.  \appendix

\section{The KK-modes}
\label{sec:modes}

The KK-modes of the gauge bosons can be obtained by solving the equation of motion
\begin{equation}
m_A^2 f_A + \left( e^{-2y} \dot f_A  \right)^\cdot = 0 \,, \label{eq:eom}
\end{equation}
where we are using the notation $\dot f\equiv df/dy$.
The $(+,+)$ boundary conditions lead to the following wavefunction
\begin{equation}
f_A^{(+,+)}(y) = C_0^{(+,+)} e^{ky} \left[ J_1(e^{ky-ky_1} \hat{m}) + C_1^{(+,+)}  Y_1(e^{ky-ky_1} \hat{m}) \right] \,, \label{eq:fApp}
\end{equation}
where
\begin{equation}
C_1^{(+,+)} = -  \frac{J_0(e^{-ky_1}\hat{m})}{Y_0(e^{-ky_1}\hat{m})}
\end{equation}
guarantees the Neumann boundary condition in the UV brane, and $J_\alpha$ and $Y_\alpha$ correspond to the Bessel functions of the first and second kind respectively. We have defined here $\hat{m} \equiv m/\rho$ with $\rho = e^{-ky_1}k$. 

By the same way, the boundary conditions $(-,+)$ lead to
\begin{equation}
f_A^{(-,+)}(y) = C_0^{(-,+)} e^{ky} \left[ J_1(e^{ky-ky_1} \hat{m}) + C_1^{(-,+)}  Y_1(e^{ky-ky_1} \hat{m}) \right] \,, \label{eq:fAmp}
\end{equation}
where
\begin{equation}
C_1^{(-,+)} = - \frac{J_1(e^{-ky_1}\hat{m})}{Y_1(e^{-ky_1}\hat{m})} 
\end{equation}
guarantees the Dirichlet boundary condition in the UV brane. In these expressions $C_0^{(+,+)}$ and $C_0^{(-,+)}$ are arbitrary constants. Notice that a constant $f_A(y)$ fulfills the $(+,+)$ boundary conditions, and from Eq.~(\ref{eq:eom}) one finds that this corresponds to a zero mode. The $(-,+)$ boundary conditions, however, do not lead to zero modes.

In the limit of large $ky_1$, the Neumann boundary conditions in the IR brane lead to the following equations for the eigenvalues
\begin{eqnarray}
0 &=& J_0(\hat{m}_{++}) + \frac{\pi}{2} Y_0(\hat{m}_{++}) \frac{1}{ky_1} + {\cal O}(1/k^2y_1^{2})   \,,  \\
0 &=& J_0(\hat{m}_{-+}) + \frac{\pi}{4} \hat{m}_{-+}^2 Y_0(\hat{m}_{-+}) e^{-2ky_1}  + {\cal O}(e^{-4ky_1})  \,,
\end{eqnarray}
for $(+,+)$ and $(-,+)$ boundary conditions respectively. Taking into account the expansion of the Bessel function $J_0(\hat{m} + \delta\hat{m}) = J_0(\hat{m}) - J_1(\hat{m}) \delta\hat{m}  + {\cal O}(\delta\hat{m}^2)$, one finds the following eigenvalues
\begin{eqnarray}
\hat{m}_{++}^{(n)} &=& \hat{m}_0^{(n)} + \frac{\pi}{2} \frac{Y_0(\hat{m}_0^{(n)})}{J_1(\hat{m}_0^{(n)})} \frac{1}{ky_1} + {\cal O}(1/k^2y_1^{2}) \,, \label{eq:m_pp}\\
\hat{m}_{-+}^{(n)} &=& \hat{m}_0^{(n)} + \frac{\pi}{4} \hat{m}_0^{(n)\,2} \frac{Y_0(\hat{m}_0^{(n)})}{J_1(\hat{m}_0^{(n)})} e^{-2ky_1} + {\cal O}(e^{-4ky_1}) \,, \label{eq:m_mp}
\end{eqnarray}
where $\hat{m}_0^{(n)}$ is the $n$-th zero of the $J_0(\hat{m})$ function, in particular: $$\hat{m}_0^{(n)} = \{2.405, 5.520, 8.654, 11.792, 14.931, \\ \cdots \}\,.$$ The second term in the right-hand side of Eq.~(\ref{eq:m_mp}) leads to corrections of ${\cal O}(10^{-28}) - {\cal O}(10^{-30})$ for the five lightest eigenvalues when $ky_1 = 35$, so that this correction can be considered negligible. The correction in Eq.~(\ref{eq:m_pp}) is $\approx 0.045 $, so the difference between the eigenvalues is then
\begin{equation}
\hat{m}_{++}^{(n)} - \hat{m}_{-+}^{(n)} = \frac{\pi}{2} \frac{Y_0(\hat{m}_0^{(n)})}{J_1(\hat{m}_0^{(n)})} \frac{1}{ky_1} + {\cal O}(1/k^2y_1^{2})
\end{equation}
of order $0.045$ for all the modes. This difference will be neglected throughout this paper.

Let us now compute the value of the coupling $ f^n_{W_R}(ky_1) = f_A^{(-,+), \, n}(ky_1)$, where we are normalizing the wave functions such that, Eq.~(\ref{eq:normalizacion}),
\begin{equation}
\int_0^{y_1} dy f_A^2(y) = y_1 \,. 
\end{equation}
The function $f_A(y)$ grows with $y$, so that this integral is dominated by the regime close to $y \simeq y_1$. In this regime the dominant contribution to the wave function is the term $\sim e^{ky} J_1(e^{ky-ky_1} \hat{m})$ in Eqs.~(\ref{eq:fApp}) and (\ref{eq:fAmp}), i.e.
\begin{equation}
f_A(y) \simeq  C_0 e^{ky} J_1(e^{ky-ky_1} \hat{m})  \,.  \label{eq:fA_approx}
\end{equation}

If we focus on the $(-+)$ solution, then one has
\begin{eqnarray}
  y_1 &\simeq& (C_0^{(-,+)})^2 \int_0^{y_1} dy \, e^{2ky}  [J_1(e^{ky-ky_1} \hat{m})]^2 \nonumber \\
&\simeq& (C_0^{(-,+)})^2 \int_0^{y_1} dy \, e^{2ky}  \left( [J_1(e^{ky-ky_1} \hat{m})]^2 + J_1(e^{ky-ky_1} \hat{m})\frac{1}{k} \frac{d}{dy} J_1(e^{ky-ky_1} \hat{m}) \right)  \nonumber \\
&=& \frac{1}{2k} (C_0^{(-,+)})^2 \int_0^{y_1} dy \frac{d}{dy} \bigg[  e^{2ky}  \left[J_1(e^{ky-ky_1} \hat{m})\right]^2  \bigg]  \nonumber \\
&\simeq& \frac{1}{2k} (C_0^{(-,+)})^2  e^{2ky_1}  [J_1(\hat{m})]^2 \,. \label{eq:normmp}
\end{eqnarray}
In the second equality we have added a term whose integral is vanishing when $\hat{m}$ is an eigenvalue of $J_0(\hat{m})$. To see this, let us note that
\begin{equation}
e^{2ky} J_1(e^{ky-ky_1} \hat{m}) \frac{d}{dy} J_1(e^{ky-ky_1} \hat{m}) = \frac{d}{dy} \bigg[ \frac{1}{2} e^{2ky}  J_0(e^{ky-ky_1} \hat{m})  J_2(e^{ky-ky_1} \hat{m})  \bigg] \,.
\end{equation}
This implies that after integrating this term in $\int_0^{y_1} dy$, the result is $\propto J_0(\hat{m})$, which is vanishing~\footnote{We are neglecting terms as  $J_\alpha(e^{-ky_1}\hat{m})$, since we consider large $ky_1$ values.}. From Eqs.~(\ref{eq:fA_approx}) and (\ref{eq:normmp}) one finally finds
\begin{equation}
|f_A^{(-+),\, n}(y_1)| \simeq \sqrt{2 ky_1} \,.  \label{eq:fAy1}
\end{equation}
This result is valid for any eigenvalue, in the approximation where we are neglecting corrections of ${\cal O}(e^{-2ky_1})$ for the lightest eigenvalues. The wave functions with boundary conditions $(++)$ have some small deviations with respect to Eq.~(\ref{eq:fAy1}) but we also find $|f_A^{(++), \, n}(y_1)| \simeq \sqrt{2ky_1}$ for the non-vanishing modes. Therefore in this paper we will use the approximation where 
\be
|f_A^{(++), \, n}(y_1)| \simeq |f_A^{(-+), \, n}(y_1)|\simeq \sqrt{2ky_1} \,.
\ee


\bibliographystyle{JHEP}
\bibliography{refs}

\end{document}